\documentclass[%
 reprint,
 amsmath,amssymb,
 aps,longbibliography,
prd
]{revtex4-1}

\usepackage{amsthm}
\usepackage{lmodern}
\usepackage{graphicx}
\usepackage{dcolumn}
\usepackage{bm}
\usepackage{hyperref}
\hypersetup{linktocpage,colorlinks,citecolor={blue},pdfdisplaydoctitle=true,pdfpagemode=UseOutlines,bookmarksnumbered=true}
\usepackage{mathrsfs,dsfont}
\usepackage{color}
\usepackage{microtype}

\definecolor{cbl}{rgb}{0,0,1}
\definecolor{darkblue}{rgb}{0.0,0.0,0.0}

\newcommand{\cbl}[1]{\textcolor{darkblue}{#1}} 

\definecolor{crd}{rgb}{0.0,0,0}

\newcommand{\bra}[1]{\langle #1 |} 
\newcommand{\ket}[1]{| #1 \rangle } 
\newcommand{\upd}{\mathrm{d}}
\newcommand{\tr}{\mathrm{tr}}
\newcommand{\ie}[0]{\textit{i.e.} }
\newcommand{\eg}[0]{\textit{e.g.} }
\newcommand{\xb}[0]{\mathbf{x}}
\newcommand{\yb}[0]{\mathbf{y}}

\newcommand{\tphi}{\widetilde{\phi}}
\newcommand\e{\mathrm{e}}
\newcommand\trans{\mathsf{T}}
\newcommand\mydots{\makebox[1em][c]{.\hfil.\hfil.}}

\newcommand\tsize{0.6}
\newcommand\msize{0.6}

\DeclareMathOperator*{\argmax}{argmax} 

\theoremstyle{remark}

\theoremstyle{definition}
\newtheorem{definition}{Definition}

\theoremstyle{plain}
\newtheorem{proposition}{Proposition}

\theoremstyle{remark}

\theoremstyle{plain}

\begin{document}
\title{Continuous Tensor Network States for Quantum Fields}
\author{Antoine Tilloy}
\email{antoine.tilloy@mpq.mpg.de}
\author{J. Ignacio Cirac}
\email{ignacio.cirac@mpq.mpg.de}
\affiliation{Max-Planck-Institut f\"ur Quantenoptik, Hans-Kopfermann-Stra{\ss}e 1, 85748 Garching, Germany}
\date{\today}

\begin{abstract}
We introduce a new class of states for bosonic quantum fields which extend tensor network states to the continuum and generalize continuous matrix product states (cMPS) to spatial dimensions~$d\geq 2$. By construction, they are Euclidean invariant, and are genuine continuum limits of discrete tensor network states. Admitting both a functional integral and an operator representation, they share the important properties of their discrete counterparts: expressiveness, invariance under gauge transformations, simple rescaling flow, and compact expressions for the $N$-point functions of local observables. While we discuss mostly the continuous tensor network states extending Projected Entangled Pair States (PEPS), we propose a generalization bearing similarities with the continuum Multi-scale Entanglement Renormalization Ansatz (cMERA).
\end{abstract}

\maketitle

\section{Introduction}

\color{darkblue}
Tensor Network States (TNS) provide an efficient parameterization of physically relevant many-body wavefunctions on the lattice \cite{cirac2009-rev,evenbly2014-rev}. Obtained from a contraction of low-rank tensors on so-called virtual indices, they economically approximate the states of systems with local interactions in thermal equilibrium. Their number of parameters scales only polynomially with the lattice size \cite{molnar2015,hastings2006}, circumventing the exponential growth of the Hilbert space dimension. TNS have led to powerful numerical methods to compute the physical properties of complex system \cite{orus2014-rev,orus2008-nonunitary_tebd,orus2012-corner}, most notably in one spatial dimension $d=1$, where Matrix Product States (MPS) \cite{fannes1992}, the simplest incarnation of TNS, are at the basis of what is arguably the most successful method to describe strongly correlated systems \cite{white1992,white1993,schollwock2005}. In higher dimensions $d\geq 2$, accurate results \cite{nielsen2017} have also been obtained using Projected Entangled-Pair states (PEPS) \cite{verstraete2004}, a natural generalization of MPS. Another family of TNS, Multi-Scale Renormalization Ansatz (MERA) \cite{vidal2007}, has proved well suited to describe scale invariant states \cite{montangero2009,evenbly2013} appearing in critical phenomena.

Beyond numerical computations, TNS provide important insights into the nature of many-body quantum systems, and have helped describe and classify their physical properties. By design, their entanglement obeys the area law \cite{srednicki1993,wolf2008,eisert2010}, which is a fundamental property of low energy states of systems with local interactions. They enable a succinct classification of symmetry protected \cite{pollmann2010,chen2011,schuch2011,chen2013} and topological phases of matter \cite{schuch2010,bultinck2017}. TNS also have a built-in bulk-boundary correspondence \cite{cirac2011}, which makes close connections to physical phenomena appearing in exotic materials \cite{laughlin1983,wen1991}. Finally, they can be used to build toy models illustrating the holographic principle and the celebrated AdS/CFT correspondence \cite{swingle2012,pastawski2015,hayden2016}.

For regular spin lattices, PEPS assign a tensor to each lattice site, with $2z$ virtual and one physical (spin) indices, where $z$ is the coordination number. The virtual indices are contracted according to the lattice geometry, yielding a wavefunction for the spin degrees of freedom. This description is particularly useful in translationally invariant systems, as this symmetry may simply be imposed by choosing the same tensor on each site. For MERA \cite{vidal2007,vidal2008}, a tree-like structure of two types of tensors is used. In both cases, the whole many-body wavefunction is determined by one or few tensors, which encode all the physical properties.

An important challenge in the theory of TNS is the generalization from lattice to continuous systems. Such an extension would allow the direct study of Quantum Field Theories, without the need for a prior breaking of spatial symmetries with a discretization. Further, the continuum provides a whole range of exact and approximate analytic techniques (such as exact Gaussian functional integrals, saddle-point approximations, or diagrammatic expansions) that have no obvious discrete counterparts and that could provide useful additions to the TNS toolbox.

A natural way to carry out such a program is to simply take the continuum limit of a TNS, by letting the lattice spacing tend to zero while appropriately rescaling the tensors. In fact, this has been done in one spatial dimension, $d=1$, where it yields continuous matrix product states (cMPS) \cite{verstraete2010,haegeman2013}. In higher dimensions, however, the task does not seem trivial. Naive extensions of cMPS have a preferred spatial direction and break Euclidean symmetries \cite{jennings2015}. In \cite{jennings2015}, a proposal for cPEPS was put forward to overcome such a limitation, but the resulting state was no longer obtained from the continuum limit of a TNS. Thus, so far there seems to be no fully satisfactory way of extending TNS to the continuum in $d\geq 2$.

In this article we propose a definition of continuous tensor network states (cTNS) that naturally extends TNS to the continuum. We obtain them as a genuine continuum limit of TNS, but manage to preserve Euclidean invariance. As in previous works \cite{jennings2015, caputa2017-prl,caputa2017-jhep}, we exploit the similarity between a tensor contraction over the indices lying on the links of a tensor network and a functional integral over a field living on the continuum limit of this mesh. The key difference lies in the way the continuum limit is taken in higher dimensions: As we shall argue, the $d=1$ case of cMPS is too peculiar to be directly extended.

The first definition of cTNS we will propose in section~\ref{sec:definitions} takes the form of a functional integral over auxiliary scalar fields as advertised. From this definition, which makes local Euclidean invariance manifest, we will derive an operator representation similar to the one used for cMPS. Importantly, we will show in section~\ref{sec:discrete} how this ansatz can be obtained from a continuum limit of a discrete TNS. We will then study some of its properties reminiscent of the discrete: its ability to approximate (possibly inefficiently) all states~(\ref{subsec:expressiveness}), its redundancy under some so called Gauge transformations~(\ref{subsec:gaugeinvariance}), which play a crucial role for PEPS, its flow under scaling transformations~(\ref{subsec:renormalization}), and its cMPS approximation in some carefully chosen limit~(\ref{subsec:recovering_cmps}). We will then propose various methods to carry computations with Gaussian and non-Gaussian cTNS (\ref{sec:computations}). While most of our approach is aimed at the continuum limit of PEPS, we will finally generalize it to MERA-like states (and more exotic TNS) in arbitrary dimensions by including a metric and restricting physical fields to a boundary~(\ref{sec:generalizations}).

\color{black}

\section{Continuous tensor network states} \label{sec:definitions}
We start by giving two equivalent definitions of continuous tensor network states, leveraging a functional integral and an operator representation. Our objective at this stage is only to provide a definition of a class of states for \cbl{bosonic} quantum fields, with only a crude intuition for why such an object could indeed be a good definition of a cTNS. We forgo the derivation of this cTNS from a class of discrete tensor networks to the following section. 

\subsection{Functional integral representation}

\subsubsection{State definition}

We begin with the functional integral representation. It will be the most direct to derive from the discrete and \cbl{makes Euclidean symmetries manifest}.
\begin{definition}[Functional integral formulation]\label{def:pathintegral}
A continuous tensor network state (cTNS) of a bosonic quantum field on a domain $\Omega \subset \mathds{R}^d$ with boundary $\partial \Omega$, is a state $\ket{V,B,\alpha}$ parameterized by 2 functions $V$ and $\alpha$: $\mathds{R}^D\rightarrow \mathds{C}$, and a boundary functional $B$: $L^2(\partial \Omega) \rightarrow \mathds{C}$ defined by the functional integral on an auxiliary $D$-component field~$\phi$:
\begin{equation}\label{eq:pathintegral}
\begin{split}
    &\ket{V,B,\alpha} = \\
    &\int\! \mathcal{D} \phi \, B(\phi|_{\partial \Omega})  \exp\bigg\{\!\!-\!\!\int_\Omega \upd^d x \; \frac{1}{2}\sum_{k=1}^D\left[\nabla \phi_k(x)\right]^2 \\
    &\hskip3.0cm+ V[\phi(x)] - \alpha[\phi(x)] \, \psi^\dagger(x)\bigg\} \ket{0},
\end{split}
\end{equation}
where $\ket{0}$ is the physical Fock vacuum state, $[\psi(x),\psi^\dagger(y)] = \delta^d(x-y)$, $\phi=[\phi_k]_{k=1}^D$. The functions $\alpha$ and $V$ may depend explicitly on position.
\end{definition}

\begin{figure}
    \centering
    \includegraphics[width=0.99\columnwidth]{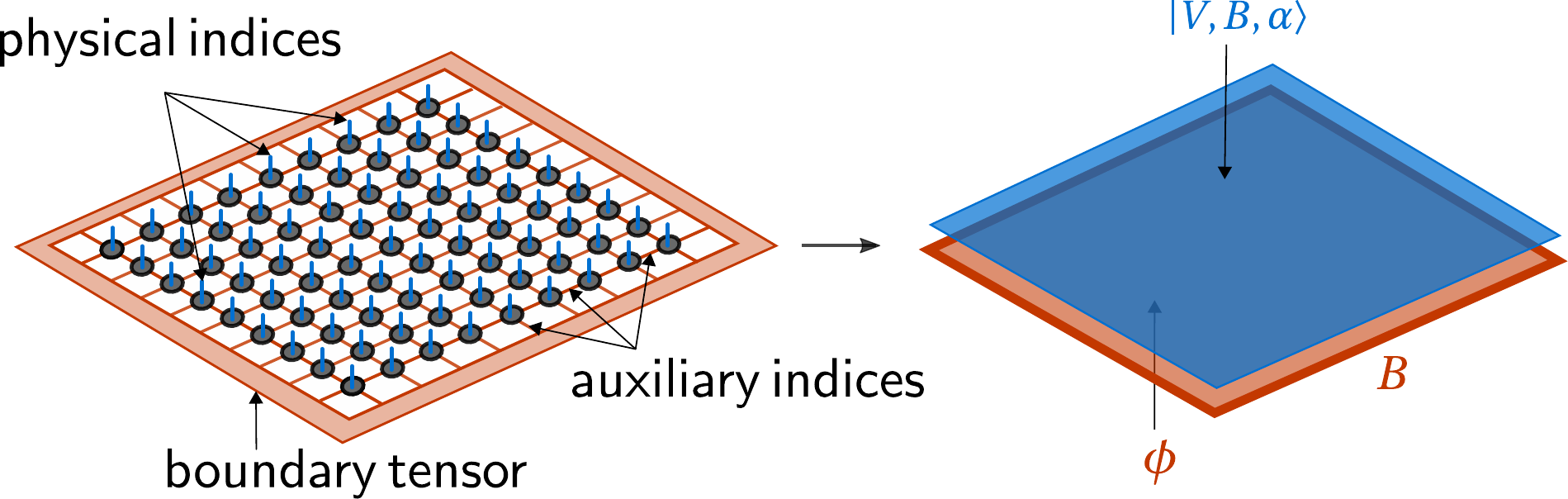}
    \caption{\textbf{Functional integral representation} -- In the discrete (left) a tensor network state is obtained from a contraction of auxiliary indices connecting the elementary tensors with each other and with a boundary tensor. In the continuum (right), the contraction is replaced by a functional integral \eqref{eq:pathintegral}, the auxiliary indices by fields $\phi$, and the boundary tensor by a boundary functional $B$. }
    \label{fig:functional}
\end{figure}

The auxiliary $D$-component field $\phi$, which is integrated over in the functional integral, is the continuous equivalent of the auxiliary bond indices that are contracted in tensor network states (see Fig. \ref{fig:functional}). This intuition will be made more precise in the next section. For this reason, we call $D$ the \emph{bond field dimension}. As we shall see in section~\ref{sec:properties}, the bond field dimension $D$ bears similarities with the bond dimension $\chi$ of discrete tensor network states.

If $\Omega$ is $\mathds{R}^d$ or a torus with periodic boundary conditions, $B$ can simply be set to $1$. \cbl{In that case, the state and its associated properties depend only on $V$ and $\alpha$, in the same way a TNS depends only on local tensors. If $V$ and $\alpha$ do not depend explicitly on $x$, the cTNS describes a translationally invariant state.} More generally, if $\partial \Omega \neq \emptyset$, the boundary functional could induce \eg:
\begin{enumerate}
    \item Dirichlet boundary conditions: $B(\phi|_{\partial \Omega}) \sim \delta(\phi|_{\partial \Omega})$ fixing $\phi|_{\partial \Omega}=0$ in the functional integral,
    \item Neumann boundary conditions: $B(\phi|_{\partial \Omega}) \sim \delta(\nabla \phi\cdot \mathbf{n})$ fixing $\nabla \phi\cdot \mathbf{n}|_{\partial \Omega}=0$ where $\mathbf{n}$ is normal to $\partial \Omega$,
    \item Something more general, given \eg by a quasi local functional:
    \begin{equation}\label{eq:extendedbc}
        B(\phi|_{\partial \Omega}) = \exp\left\{-\oint_{\partial \Omega} \upd^{d-1} x \; \mathcal{L}\left[\phi(x),\nabla\phi(x)\right]\right\}
    \end{equation}
    where $\mathcal{L}$ is a function from $\mathds{R}^{(d+1)D}$ to $\mathds{C}$. This latter option will be generated naturally when we discuss gauge invariance in \ref{subsec:gaugeinvariance}.
\end{enumerate}

We may rewrite expression \eqref{eq:pathintegral} more explicitly as a sum over unnormalized field coherent states. Introducing the massless free field probability measure $\upd \mu(\phi)$ for the auxiliary field:
\begin{equation}
    \upd \mu(\phi) = \mathcal{D} \phi \exp\left[-\frac{1}{2}\int_\Omega \upd^d x \; \sum_{k=1}^D\left[\nabla \phi_k(x)\right]^2\right],
\end{equation}
 and a complex amplitude $\mathcal{A}_V(\phi)$:
 \begin{equation}
     \mathcal{A}_V(\phi)= B(\phi|_{\partial \Omega})\exp\left\{-\int_\Omega \upd^d x \;  V[\phi(x)] \right\}
\end{equation}
yields:
\begin{equation}
    \ket{V,B,\alpha} = \int \upd \mu(\phi) \, \mathcal{A}_V(\phi)\,  \ket{\alpha(\phi)},
\end{equation}
where  $\ket{\alpha(\phi)}=\exp\left\{\int_\Omega \upd^d x \; \alpha[\phi(x)]\,\psi^\dagger(x)\right\} \ket{0}$ is an unnormalized field coherent state. Hence, just like cMPS in dimension $1$, cTNS are a generalization of field coherent states. \cbl{The latter are obtained \eg if $\upd \mu(\phi)$ is only non-zero for a given $\phi$ (for an infinitely deep $V$), or, in the homogeneous case, if $\alpha$ is constant.}

\subsubsection{N-particle Wave function}
A generic state $\ket{\Psi}$ in the bosonic Fock space $\mathcal{F}[L^2(\mathds{R}^d,\mathds{C})]$ can be expanded into a sum of $n$ particle wave functions $\varphi_n$:
\begin{equation}\label{eq:wavefunctionrepresentation}
\ket{\Psi} = \sum_{n=0}^{+\infty} \int_{\Omega^n}\!\!\!\!\upd x_1 \mydots \upd x_n \frac{\varphi_n(x_1,\mydots,x_n)}{n!} \; \psi^\dagger(x_1)\mydots \psi^\dagger(x_n) \ket{0}
\end{equation}
where $\varphi_n$ is a completely symmetric function of its coordinates.
Simply expanding the exponential of eq. \eqref{eq:pathintegral} gives, for the cTNS $\ket{V,B,\alpha}$:
\begin{equation}\label{eq:wave-path}
    \varphi_n(x_1,\cdots,x_n) = \! \int \! \upd \mu(\phi) \mathcal{A}_V(\phi)\,  \alpha[\phi(x_1)] \cdots \alpha[\phi(x_n)]
\end{equation}
It provides an equivalent definition of the cTNS.

\subsubsection{Correlation functions}
A state is also fully characterized by its (equal time) correlation functions. To compute them, we first introduce the generating functionals for real sources $j',j$:
\begin{align}
\mathcal{Z}_{j',j}&=\frac{\bra{V,B,\alpha} \exp\left(\int_\Omega j'\cdot \psi^\dagger \right) \exp\left(\int_\Omega j\cdot \psi  \right) \ket{V,B,\alpha}}{\langle V,B,\alpha|V,B,\alpha\rangle} \nonumber\\
\widetilde{\mathcal{Z}}_{j',j}&=\frac{\bra{V,B,\alpha} \exp\left(\int_\Omega j\cdot \psi \right) \exp\left(\int_\Omega j'\cdot \psi^\dagger  \right) \ket{V,B,\alpha}}{\langle V,B,\alpha|V,B,\alpha\rangle}\label{eq:ordered-def}
\end{align}
They generate the normal ordered and anti normal ordered correlation functions respectively. For example, it is straightforward to verify that:
\begin{align}\label{eq:correlexample}
\langle\psi^\dagger(x) \psi(y)\rangle&:= \frac{\langle V,B,\alpha | \psi^\dagger(x) \psi(y)| V,B,\alpha\rangle}{\langle V,B,\alpha | V,B,\alpha\rangle}\\
&= \frac{\delta}{\delta j'(x)}\frac{\delta}{\delta j(y)} \mathcal{Z}_{j',j} \bigg|_{j,j'=0}.
\end{align}
Using the formula for the overlap of (unnormalized) field coherent states,
\begin{equation}
\langle \beta |\alpha\rangle = \exp\left(\int_{\Omega}\upd x\, \beta^*(x)\,\alpha(x)\right)
\end{equation}
and writing $\mathcal{N}=\langle V,B,\alpha|V,B,\alpha\rangle$ we get:
\begin{widetext}
\begin{align}\label{eq:antiordered}
\widetilde{\mathcal{Z}}_{j',j}&= \frac{1}{\mathcal{N}}\int\!\upd \mu(\phi') \upd \mu (\phi) B(\phi) B^*(\phi')\,\exp\left\{\!-\!\!\int_\Omega \,V^*[\phi'] + V[\phi]-\left(\alpha^*[\phi']+j\right)\cdot\left(\alpha[\phi]+j'\right)\right\}\!.
\end{align}
We observe an important fact which is that the function $\alpha$ appears squared, hence an $\alpha$ quadratic in the field already brings non-Gaussianities.
To compute $\mathcal{Z}_{j',j}$, one applies the Baker-Campbell-Hausdorff (BCH) formula to \eqref{eq:ordered-def} to push annihilation operators to the right and get back to a computation of field coherent state overlaps. We obtain:
\begin{equation}\label{eq:ordered}
\begin{split}
&\mathcal{Z}_{j',j}= \frac{1}{\mathcal{N}} \int\!\upd \mu(\phi') \upd \mu (\phi)B(\phi) B^*(\phi')\,\exp\bigg\{ \!-\!\!\int_\Omega \,V^*[\phi'(x)] + V[\phi] -\alpha^*[\phi']\cdot\alpha[\phi]-j\cdot\alpha[\phi]-j'\cdot \alpha^*[\phi']\bigg\},
\end{split}
\end{equation}
hence the same as \eqref{eq:antiordered} but for the removal of the product $j\cdot j'$, which is responsible for the divergent equal point contributions upon functional differentiation.
\end{widetext}

\subsection{Operator representation}

\subsubsection{State definition}
We now provide an equivalent operator representation of cTNS. For simplicity we restrict ourselves to domains of $\mathds{R}^d$ that can be split into Cartesian products $\Omega=[-T/2,T/2] \times S$ where $\partial S = \emptyset$. This is not strictly necessary but substantially simplifies the definition.

\begin{figure}
    \centering
    \includegraphics[width=0.99\columnwidth]{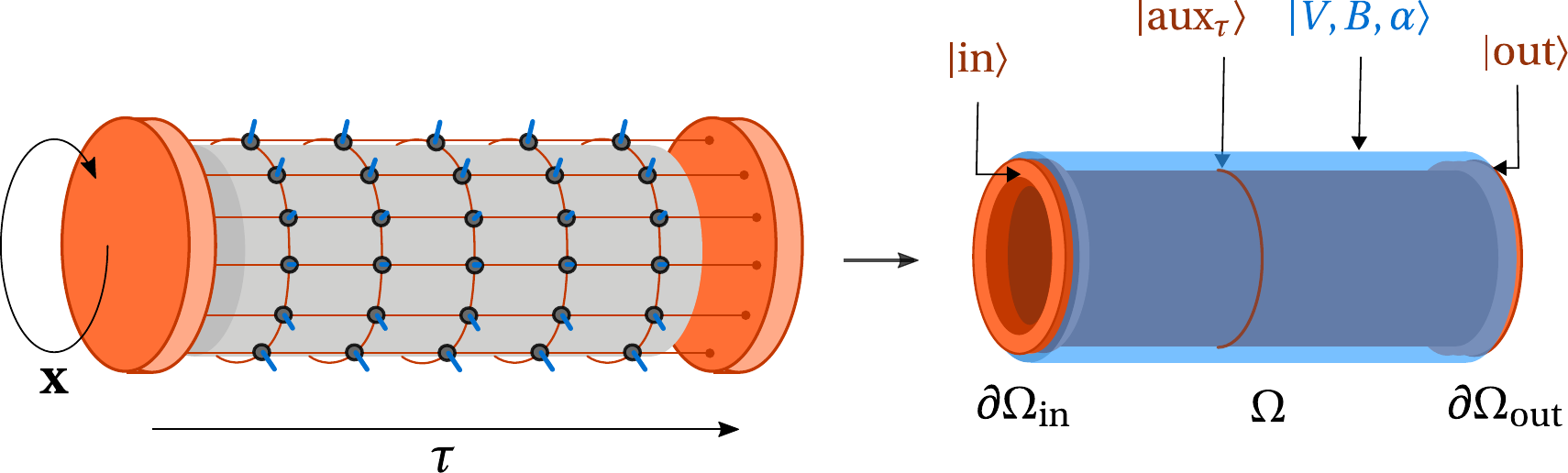}
    \caption{\textbf{Operator representation} -- Tensor networks in $d \geq 2$ can be defined through an auxiliary dynamics in $d-1$ dimensions. In the continuum, the physical $d$ dimensional quantum field $\ket{V,B,\alpha}$ is obtained through a joint non-unitary evolution (eq. \eqref{eq:operator}) with an auxiliary $d-1$ dimensional quantum field $\ket{\text{aux}_\tau}$.}
    \label{fig:operator}
\end{figure}

\begin{definition}[Alternative operator formulation]\label{def:operator}
For a domain $\Omega$ that can be written as a cartesian product $\Omega=[-T/2,T/2] \times S$, we write $x=(\tau,\xb)$ where $\tau \in [-T/2,T/2]$ and $\xb \in S$. A cTNS is then defined as:
\begin{equation}\label{eq:operator}
\begin{split}
\ket{V,B,\alpha}=\tr \bigg[\hat{B} \mathcal{T} \exp\bigg(\!\!-\!\!\!\! \int\limits_{-T/2}^{T/2} \!\!\!\upd \tau \!\int_S \!\upd\xb\;\sum_{k=1}^D \frac{\left[\hat{\pi}_k(\xb)\right]^2 }{2} \\
+  \frac{[\nabla \hat{\phi}_k(\xb)]^2}{2}  + V[\hat{\phi}(\xb)] - \alpha[\hat{\phi}(\xb)]\,\psi^\dagger(\tau,\xb) \bigg)\bigg]\ket{0}
\end{split}
\end{equation}
where $\mathcal{T}$ is the $\tau$-ordering operator, $\hat{\phi}_k(\xb)$ and $\hat{\pi}_k(\xb)$ are $k$ independent canonically conjugated pairs of (auxiliary) field \emph{operators}: $[\hat{\phi}_k(\xb),\hat{\phi}_l(\yb)]=0$, $[\hat{\pi}(\xb)_k,\hat{\pi}_l(\yb)]=0$, and $[\hat{\phi}_k(\xb),\hat{\pi}_l(\yb)]=i \delta_{k,l}\,\delta^{d-1}(\xb-\yb)$.  These operators act on $\mathscr{H}_\text{aux} =\mathcal{F}[L^2(S)]^D $, \ie $D$ copies of a bosonic Fock space on a $d-1$ dimensional space. The trace is taken over this auxiliary Hilbert space. As before, $V$ and $\alpha$ may depend on $\xb$ and $\tau$.
\end{definition}
The operator $\hat{B}$ acts on $\mathscr{H}_\text{aux}$ and fixes the boundary conditions, \eg $\hat{B}=\mathds{1}$ encodes periodic boundary conditions on the coordinate $\tau$. Another natural option is to take $\hat{B} = \ket{\text{in}}\bra{\text{out}}$, which corresponds to the situation of Fig. \ref{fig:operator}.

Definitions \ref{def:pathintegral} and \ref{def:operator} are equivalent for this subclass of domains. The proof is straightforward. One just applies the techniques of standard QFT textbooks to go from operator to functional integral representations with $\tau=it$ (see \eg \cite{peskin1995,zinn2010path}). Mainly, one discretizes the $\tau$-ordered product in \eqref{eq:operator} into a finite product of terms.
One then inserts resolutions of the identity in the field basis $\ket{\phi}$ at every time step $\Delta \tau$ and writes each resulting overlap in the conjugate momentum basis $\ket{\pi}$. Going back to the continuum limit yields a phase space functional integral which reduces to the formula of equation \eqref{eq:pathintegral} upon Gaussian integration of the conjugate momenta $\pi$. 

The boundary operator of \eqref{eq:operator} is related to the boundary functional of \eqref{eq:pathintegral} by:
\begin{equation}
    B(\phi) =\bra{\phi_\text{in}} \hat{B} \ket{\phi_\text{out}}
\end{equation}
where $\ket{\phi_\text{out}}$ and $\ket{\phi_\text{in}}$ are eigenstates of the auxiliary field operators $\hat{\phi}(\xb)$. The auxiliary field $\phi$ decomposes into $\phi = \phi_\text{in} + \phi_\text{out}$ where $\phi_\text{in}$ (resp. $\phi_\text{out}$) has support on $\partial \Omega_\text{in}$ (resp. $\partial \Omega_\text{out}$) with $\partial \Omega = \partial\Omega_\text{in} \cup \partial \Omega_\text{out}$.

As before we may propose a repackaging of formula \eqref{eq:operator}. Introducing the Hamiltonian density $\mathcal{H}(\xb)=\mathcal{H}_0(\xb) + V[\hat{\phi}(\xb)]$ with $\mathcal{H}_0(\xb)=\sum_{k=1}^D \frac{[\hat{\pi}_k(\xb)]^2 + [\nabla\hat{\phi}_k(\xb)]^2}{2}$ yields:
\begin{equation}
\begin{split}
\ket{V,B,\alpha}=\tr \bigg[\hat{B} \mathcal{T} \exp\Big(-&\int_{-T/2}^{T/2} \!\!\!\upd \tau \!\int_S \!\upd\xb \; \mathcal{H}(\xb)\\
-& \alpha[\hat{\phi}(\xb)]\,\psi^\dagger(\tau,\xb) \Big)\bigg]\ket{0}.
\end{split}
\end{equation}
This is a straightforward extension of the cMPS definition \cite{verstraete2010} (recalled in \ref{subsec:recovering_cmps}) with $\hat{Q}\sim -\mathcal{H}(\xb)$ and $\hat{R}\sim \alpha[\hat{\phi}(\xb)]$.

\subsubsection{N-particle wave function}
The N-particle wave-function $\varphi_n$ defined in \eqref{eq:wavefunctionrepresentation} can also be computed in the operator representation. For $-T/2< \tau_1<\cdots < \tau_n <T/2$, we get, expanding the $\tau$-ordered exponential \eqref{eq:operator} into an infinite product:
\begin{equation}\label{eq:wave-operator}
 \varphi_n=   \tr\left[\hat{B}\; \hat{G}_{T,\tau_n}\, \hat{\alpha}(x_n)  \, \hat{G}_{\tau_n,\tau_{n-1}} \, \hat{\alpha}(x_{n-1}) \cdots \hat{\alpha}(x_1)\,  \hat{G}_{\tau_1,0} \right]
\end{equation}
with $\hat{\alpha}(x_n) = \alpha[\hat{\phi}(\xb_n)]$, $\hat{G}_{u,v} = \mathcal{T}\exp[-\int_v^u \upd \tau \!\int_S \upd \xb \mathcal{H}(\xb)]$. As for cMPS \cite{verstraete2010}, we may interpret $\hat{G}$ as a propagator and $\hat{\alpha}$ as a scattering matrix creating a particle. \cbl{It is the very specific form of $\mathcal{H}$ and hence of $\hat{G}$ that is responsible for the Euclidean symmetries of the resulting state. Generalizing cMPS starting directly from \eqref{eq:wave-operator} would make it hard to guess an appropriate expression for $\hat{G}$.}

Note that \eqref{eq:wave-operator} amounts to taking as wave function a correlation function of auxiliary quantum fields. This is similar in spirit with the Moore-Read states \cite{moore1991} used for Hall physics or with the infinite matrix product states \cite{cirac2009} used for critical spin chains.

\subsubsection{Correlation functions}

Finally, we may provide an expression for the generating functionals $\mathcal{Z}_{j',j}$  $\widetilde{\mathcal{Z}}_{j',j}$ in the operator representation. Exploiting the operator definition of the cTNS \eqref{eq:operator}, expanding the $\tau$-ordered exponential into an infinite product of infinitesimal exponentials, we get:
\begin{equation}\label{eq:operator-antiordered}
    \widetilde{\mathcal{Z}}_{j',j} = \tr\left[B\otimes B^* \, \mathcal{T}\exp\left(\int_{-T/2}^{T/2} \mathds{T}_{j'j}\right)\right]
\end{equation}
with the \emph{transfer matrix} (with sources):
\begin{equation}
    \mathds{T}_{j'j} = \int_S -\mathcal{H}\otimes \mathds{1} -\mathds{1}\otimes \mathcal{H}^* + (\alpha[\hat{\phi}] + j') \otimes (\alpha[\hat{\phi}]^* + j)
\end{equation}
Using as before the BCH formula yields:
\begin{equation}\label{eq:operator-ordered}
        \mathcal{Z}_{j'j} = \tr\left[B\otimes B^* \mathcal{T}\exp\left\{\int_{-T/2}^{T/2} \left(\mathds{T}_{j'j} - \int_{S} j\cdot j'\right)\right\}\right]
\end{equation}
The functional derivatives can then be carried explicitly and one obtains \eg , for $-T/2<\tau_2 <\tau_1<T/2$:
\begin{equation}
\begin{split}
    \langle \psi^\dagger(x_1) \psi(x_2)\rangle = \tr &\big\{B\otimes B^* \cdot \mathcal{M}_{T/2,\tau_1} \cdot  [\mathds{1} \otimes \hat{\alpha}^*(x_1)] \\
    &\!\! \cdot \mathcal{M}_{\tau_1,\tau_2} \cdot [\hat{\alpha}(x_2)\otimes \mathds{1}] \cdot \mathcal{M}_{\tau_2,-T/2}  \big\}
\end{split}
\end{equation}
with  the map $\mathcal{M}_{u,v}= \mathcal{T}\exp[\int_v^u \mathds{T}]$ and the \emph{transfer matrix} $\mathds{T}:=\mathds{T}_{00}$. More generally, correlation functions are given by the trace of a succession of propagators $\mathcal{M}$ followed by operator insertions of $\alpha \otimes \mathds{1}$ (respectively $\mathds{1}\otimes \alpha^*$) in the positions corresponding to $\psi$ (respectively $\psi^\dagger$).

\section{Link with discrete tensor network states}\label{sec:discrete}

\subsection{(Discrete) tensor network states}
We start with a very brief reminder on tensor network states (TNS), recalling only their elementary definition. For an understanding of their efficiency in representing quantum systems of physical interest, we direct the reader to the relevant literature (\eg \cite{verstraete2008,bridgeman2017} and references therein).

TNS are variational ansatz for many-body wave functions that take the form of a contraction of local tensors. The simplest example, in spatial dimension $d=1$, is provided by matrix product states (MPS). For a translation invariant quantum spin $1/2$ chain with $N$ sites, a generic state reads:
\begin{equation}
    \ket{\psi} = \sum_{i_1,\cdots,i_n =\{-1,1\}^N} c_{i_1,\cdots,i_n}\; \ket{i_1}\otimes\cdots\otimes\ket{i_N}
\end{equation}
where the wave function $c_{i_1,\cdots,i_N}$ contains $2^N$ complex parameters. A MPS is a economical ansatz for this wave function:
\begin{equation}\label{eq:defmps}
    c_{i_1,\cdots,i_n} = \tr \left[A_{i_1}\cdots A_{i_N}\right]
\end{equation}
where $A_{-1}$ and $A_{1}$ are two $\chi\times\chi$ matrices. These matrices contain the parameters which allow to vary the state. Their size $\chi$, called the bond-dimension, encodes the depth of the variational class and upper bounds the amount of spatial entanglement that can be carried by the state.

The two matrices can be be collected into a 3-index tensor $[A_i^{k,\ell}]^{k,\ell=1\cdots \chi}_{i=\pm 1}$ written graphically:
\begin{equation}
   A_i^{k,\ell} =k\;  \begin{array}[b]{c}
   i\\
   \hbox{\includegraphics[scale=\msize]{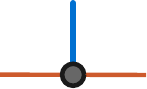}}
   \end{array}\; \ell
\end{equation}
where $i$ is the physical index and $k,\ell$ are so called bond indices. Graphically, the corresponding wave function $c$ can be written:
\begin{equation}
    c_{i_1,\cdots,i_N}=\vcenter{\hbox{\includegraphics[scale=\msize]{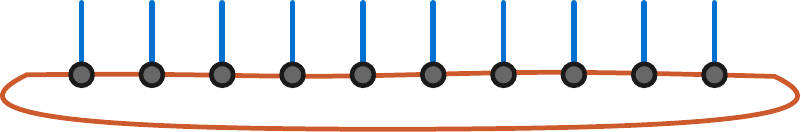}}}
\end{equation}
where joint legs of the tensor $A$ denote a summation on the corresponding index, associated with the matrix multiplication and subsequent trace in \eqref{eq:defmps}. This representation makes it natural to generalize matrix product states to \cbl{projected entangled pair states (PEPS)} \cite{verstraete2004} in arbitrary dimensions, \eg in $d=2$ for $N\times N$ sites:
\begin{equation}
 c_{i_1,\cdots,i_{N^2}}  = \vcenter{\hbox{\includegraphics[scale=\tsize]{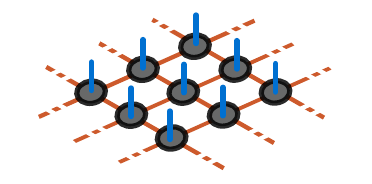}}} 
\end{equation}
This latter object, in general $d$, is what the cTNS defined in \eqref{eq:pathintegral} or \eqref{eq:operator} aims to extend to the continuum.
\subsection{Constructing cTNS}

\begin{figure}
    \centering
    \includegraphics[width=0.99\columnwidth]{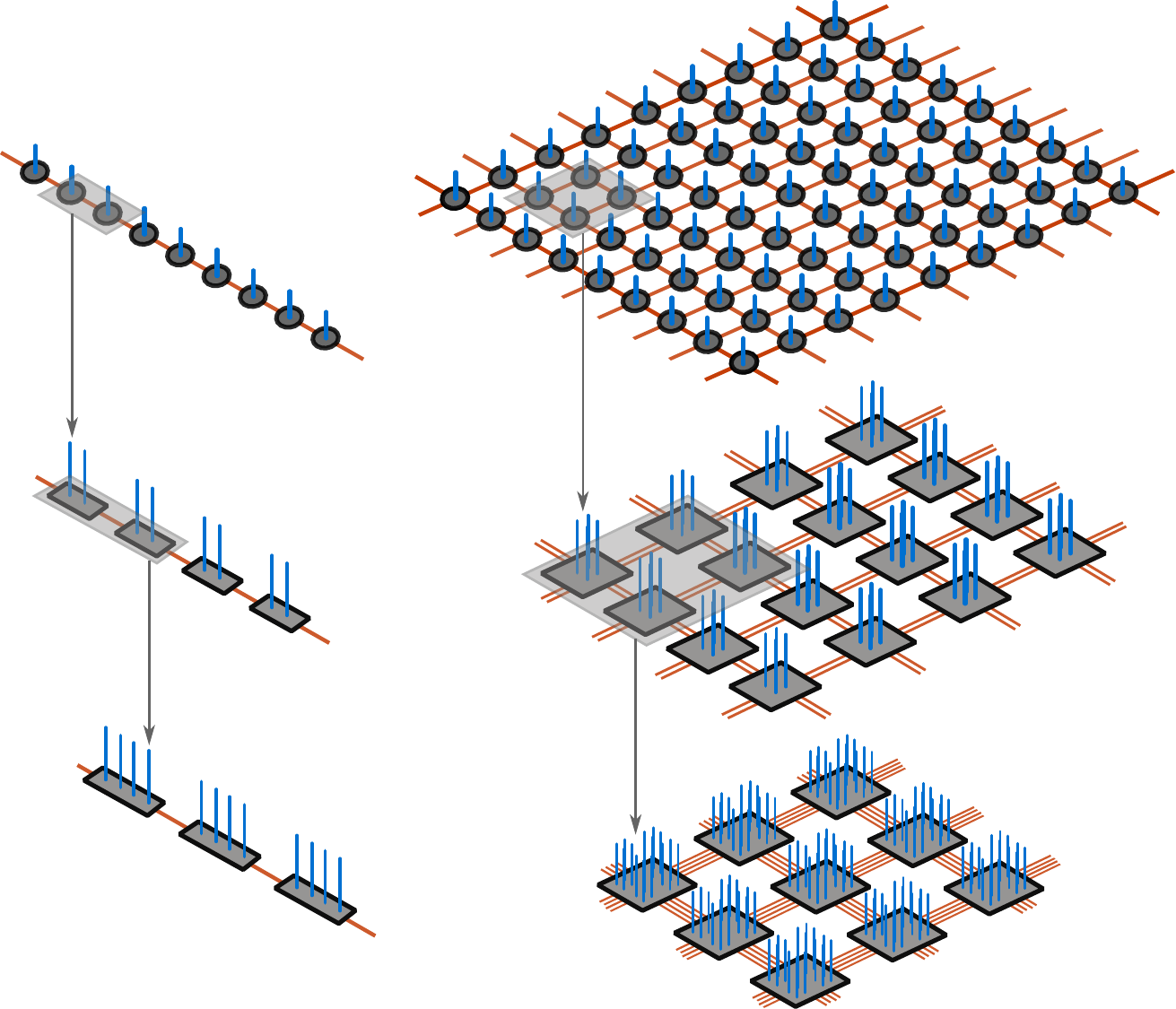}
    \caption{\textbf{Tensor blocking} -- In $d=1$, blocking does not increase the bond dimension. In $d=2$, going from the UV to the IR doubles the bond dimension at each blocking. Hence flowing the other way, from IR to UV, one reaches a trivial bond dimension after a finite number of iterations unless the initial bond dimension is infinite.}
    \label{fig:peps_renormalization}
\end{figure}

Our objective is to show how cTNS can be obtained from a limit of a discrete TNS. To motivate this discrete ansatz, we first provide heuristics for why its main characteristics seem to be required. Mainly, we aim to justify:
\begin{enumerate}
    \item why infinite bond dimension is needed, and
    \item why the trivial tensor around which we expand is of the form we postulate.
\end{enumerate}
The first point is a scaling argument. We ask for a strong notion of continuum limit: we require the discrete tensor to be approximately stable by fine graining to the UV. Namely, the discrete ansatz needs to be (at least approximately) expressible as a contraction of tensors with the same form but different parameters. Each blocking multiplies the physical dimension by $2^d$, and this is why in the continuum limit one obtains a field theory on the physical degrees of freedom. But each blocking also multiplies the bond dimension by $2^{d-1}$ (see Fig. \ref{fig:peps_renormalization}). Hence, for $d>1$, the bond dimension is increased when zooming out and decreased when zooming in. The only way to make the class of states considered approximately stable is for the bond dimension to be infinite. Notice in this respect that the $d=1$ case allows finite bond dimensions even in the continuum limit \cite{verstraete2010,haegeman2013}. It will be important to see if this peculiarity can be recovered in some appropriate limit in section \ref{subsec:recovering_cmps}. 

Note that our argument in favor of infinite bond dimension does not imply that a discrete tensor network state with finite bond dimension could not behave, at distances sufficiently large compared to the lattice spacing, like a cTNS. Rather, our argument shows that any simple space discretization of a cTNS in $d\geq 2$ into a TNS will have infinite bond dimension, even for arbitrarily small lattice spacing. As in \cite{jennings2015}, it could also be that a proper choice of boundary conditions would constrain the tensor contraction on a finite dimensional subspace, despite an apparent infinite bond dimension.

We now need to discuss more precisely the form of the elementary tensor. For notational simplicity, we now assume that an elementary tensor $\hat{T}$ to be contracted:
\begin{equation}
    \hat{T} := \vcenter{\hbox{\includegraphics[scale=\tsize]{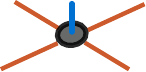}}}
\end{equation}
is a \emph{vector} in its bond indices but an \emph{operator} acting on the vacuum in the physical space, namely:
\begin{align}
\ket{\text{physical state}}&= \text{contraction}\left\{\text{network of }\hat{T}\right\}\; \ket{0}\\
&=\vcenter{\hbox{\includegraphics[scale=\tsize]{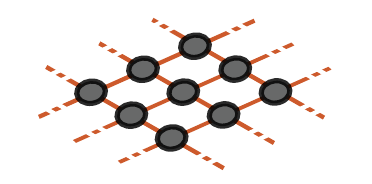}}}\,\ket{0}
\end{align}
To obtain a continuum limit, the crucial choice lies in the elementary ``trivial tensor'', acting \cbl{as the identity} on the vacuum, and around which to expand:
\begin{equation}
    \hat{T}=     \hat{T}^{(0)}  + \varepsilon^d \times \text{corrections}.
\end{equation}
Indeed, it is natural to want the tensor corresponding to zero particle in an elementary cell of the physical space to dominate. It seems that any other choice would preclude the existence of a continuum limit.
In $d=1$ dimension, there is only one natural option which is to take the tensor corresponding to the identity:
\begin{equation}
    \hat{T}^{(0)} = \vcenter{\hbox{\includegraphics[scale=\tsize]{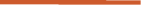}}}
\end{equation}

But in the same way as for the bond dimension, the situation is a little too trivial in $d=1$ to give a precise hint for higher dimensions. In $d>1$, there are several seemingly natural options which we have to inspect. We will discuss the $d=2$ case but the reasoning holds for any $d\geq 2$. We do not aim to prove that the tensor we will ultimately expand around is the only option, but rather that other seemingly simpler options bring difficulties.
\begin{enumerate}
\item A naive option is to generalize the identity on the auxiliary bond space by taking $\hat{T}^{(0)}_{ijkl}=\delta_{ijkl}$, that is to take an elementary tensor corresponding to a \cbl{Greenberger-Horne-Zeilinger} (GHZ) state on the bond indices:
\begin{equation}\label{eq:ghz}
\hat{T}^{(0)}=\vcenter{\hbox{\includegraphics[scale=\tsize]{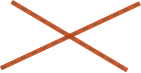}}}
\end{equation} 
As will later be manifest, this choice is too brutal and would yield a state with a trivial spatial structure.
\item Another simple option is to take the identity along a diagonal, \eg:
\begin{equation}
\hat{T}^{(0)}=\vcenter{\hbox{\includegraphics[scale=\tsize]{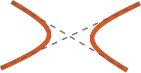}}}
\end{equation}
This is the choice that stays the closest in spirit with the $d=1$ case. The problem of such a choice (made \eg in \cite{jennings2015}) is that it picks a prefered a direction and thus makes Euclidean invariance impossible to obtain directly (that is, without analytic continuation).
\item We may combine the $2^{d-1}$ identity operator along diagonals in a sum, as an attempt to recover the Euclidean invariance lost with the previous choice:
\begin{equation}
\hat{T}^{(0)}= \vcenter{\hbox{\includegraphics[scale=\tsize]{idpepsu.pdf}}} + \vcenter{\hbox{\includegraphics[scale=\tsize]{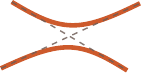}}}
\end{equation}
An issue is then that the corresponding tensor contraction contains loops:
\begin{equation}
    \vcenter{\hbox{\includegraphics[scale=\tsize]{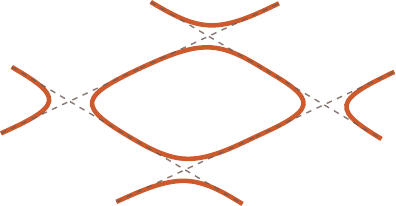}}}
\end{equation}
which yield divergent terms as $\tr [\mathds{1}]=+\infty$ for infinite bond dimension. 
\end{enumerate}
None of the natural options seems to provide a simple Euclidean invariant continuum limit. Our proposal will consist in taking a regularized version of the first possibility, in the form of a ``soft'' delta:
\begin{equation}
\hat{T}^{(0)}=\vcenter{\hbox{\includegraphics[scale=\tsize]{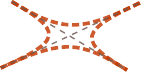}}}~~~~.
\end{equation}

\subsection{Discrete Ansatz in \texorpdfstring{$d=2$}{d=2}}

The first lesson from the previous sections is that infinite bond dimensions seem to be required. We thus write the bond indices of the elementary tensor as $D$ real numbers. In $d=2$, this means that an elementary tensor has $4$ bond indices $\phi(1)$, $\phi(2)$, $\phi(3)$, and $\phi(4) \in \mathds{R}^{D}$:
\begin{equation}\label{eq:discretedef}
\hat{T}_{\phi(1),\phi(2),\phi(3),\phi(4)}= { \scriptsize\begin{array}{ccc}
 \phi(2)\!\!\!\!\!  &  & \!\! \phi(3) \\
     &\vcenter{\hbox{\includegraphics[scale=\tsize]{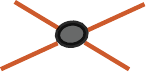}}} & \\
    \phi(1)\!\!  &  & \!\!\!\!\! \phi(4)  
\end{array}}
\end{equation}
As we mentioned before, the heart of the problem of the continuum limit lies in defining the proper trivial tensor around which to expand. We choose a ``soft'' delta:
\begin{align}
\hat{T}^{(0)}&=\vcenter{\hbox{\includegraphics[scale=\tsize]{softghz.pdf}}}\nonumber \\
&=\exp\bigg\{\frac{-1}{2}\sum_{k=1}^D[\phi_k(1)-\phi_k(2)]^2 +
[\phi_k(1)-\phi_k(4)]^2 \nonumber\\
&\hskip1cm+[\phi_k(3)-\phi_k(2)]^2+[\phi_k(3)-\phi_k(4)]^2\bigg\}\label{eq:discrete}
\end{align}
This ansatz forces the bond indices to remain close to each other, and contributes to the generation of the gradient squared term in the action. Being Gaussian, we also naturally expect its form to be stable. To this ``trivial'' part, we add local corrections of order $\varepsilon^d=\varepsilon^2$:
\begin{equation}\label{eq:defT0}
\hat{T} = \hat{T}^{(0)}\exp\left[-\varepsilon^2 V(\phi) \,\mathds{1} + \varepsilon^2 \alpha(\phi) \;\psi^\dagger(x)\right].
\end{equation}
In this expression, $\phi$ denotes whatever combination of the bond field indices $\phi(1)$, $\phi(2)$, $\phi(3)$, and $\phi(4)$. The simplest possibility is to take $\phi$ as the average of the bond indices but it does not matter for the continuum limit. The operator $\psi^\dagger(x)$ anticipates the continuum and has commutation relations $[\psi(x),\psi^\dagger(y)]=\frac{1}{\varepsilon^2} \delta_{x,y} \simeq \delta^2(x-y)$.

Ignoring the boundary conditions for now, the contraction of the tensors amounts to integrate over all the bond indices:
\begin{equation}
\ket{V,\alpha}=\!\!\int\!\!\!\! \prod_{x\in\text{lattice}}  \!\!\!\!\! \hat{T}(x) \! \prod_{k=1}^D\upd \phi_k \! \left(x+\frac{\varepsilon}{2} \mathbf{e}_1\right) \upd \phi_k\!\left(x+\frac{\varepsilon}{2} \mathbf{e}_2\right)  \ket{0}
\end{equation}
where $\mathbf{e}_1$ and $\mathbf{e}_2$ are unit vectors along the two lattice directions and the bond fields are indexed by the points on the links of the lattice where they sit. Writing $u=\frac{x^1+x^2}{\sqrt{2}}$ and $v=\frac{x^1-x^2}{\sqrt{2}}$ we see that the differences in eq. \eqref{eq:defT0} yield:
\begin{equation}
\begin{split}
\prod_{x\in\text{lattice}}\!\!\!\!  \hat{T}(x) \underset{\varepsilon\rightarrow 0}{\simeq} \exp\bigg\{-&\int \!\upd^2 x  \,  \frac{[\partial_u\phi_k(x)]^2 + [\partial_v\phi_k(x)]^2}{2}\\
&+V[\phi(x)] -\alpha[\phi(x)]\,\psi^\dagger(x)\bigg\}.
\end{split}
\end{equation}
We recognize the (rotation invariant) gradient square term of the continuum definition \ref{def:pathintegral}. Defining the path integral ``measure'' as:
\begin{equation}
\mathcal{D}\phi :\simeq \lim_{\varepsilon\rightarrow 0} \prod_{\underset{k=1\cdots D}{x\in\text{lattice}}}\upd \phi_k\left(x+\frac{\varepsilon}{2} \mathbf{e}_1\right) \upd \phi_k\left(x+\frac{\varepsilon}{2} \mathbf{e}_2\right)
\end{equation}
finaly yields the continuous tensor network state of equation \eqref{eq:pathintegral} up to boundary conditions. To get a state on the physical Hilbert space, the auxiliary fields on the boundary just have to be contracted (or integrated) against a boundary functional which we wrote $B$ in \eqref{eq:pathintegral}.

\subsection{Discrete Ansatz in general \texorpdfstring{$d$}{d}}
For $d\geq 3$ the derivation is carried along the same way as before. We just note that there is a small peculiarity in the $d=2$ case because the auxiliary fields are adimensional. To generalize equation \eqref{eq:discrete} to higher spatial dimensions $d>2$, one naturally extends the prescription of summing all the differences of the squares of the nearest bond indices $\phi(1),\cdots \phi(2 d)$. But, importantly, to obtain the continuum limit, one needs to multiply this expression by $\varepsilon^{d-2}$ where $\varepsilon$ is the length of the unit cell. 
\begin{equation}\label{eq:d-ansatz}
\hat{T}^{(0)}_{\phi(1) \cdots \phi(2d)} = \exp\left\{-\frac{\varepsilon^{d-2}}{2} \sum_{k=1}^D \cdots\right\}
\end{equation}
to obtain the integral of a gradient squared in the continuum limit.
In $d=2$, the $\varepsilon^d$ of the integration measure and $\varepsilon^{-2}$ from the gradient square cancel each other and this scaling factor does not appear.

In retrospect, it is clear why deriving the continuum limit by perturbing around the GHZ tensor \eqref{eq:ghz} would have given a trivial continuum. It would have corresponded to putting an infinitely large constant instead of $\varepsilon^{d-2}$ in \eqref{eq:d-ansatz}, an infinite ``rigidity'' that could not be compensated by locally small terms.

\section{Properties}\label{sec:properties}

We now explore the properties of cTNS that are analogous to those of their discrete counterparts.

\subsection{Stability and expressiveness}\label{subsec:expressiveness}
It is first natural to wonder how ``big'' the class of cTNS is. It could be, for example, that even for arbitrary large $D$ and arbitrary $V$, $B$ and $\alpha$, cTNS only spanned a small sector of the Fock space. As in the discrete, can any state be approximated, even if inefficiently, by a cTNS?

Let us first consider the \emph{stability} of the cTNS class.  The sum of two cTNS is still a cTNS, provided we are willing to accept singular potentials $V$. More precisely, let $\ket{V_1,B_1,\alpha_1}$ and $\ket{V_2,B_2,\alpha_2}$ be two cTNS with bond-field dimension $D_1$ and $D_2$. Then we can easily rewrite their sum as a cTNS with bond-field dimension $D_1+D_2+1$ (although it may in general require fewer auxiliary fields). For example, defining the cTNS $\ket{W_\Lambda,C,\beta}$ with:
\begin{align}
    W_\Lambda (\phi_1,\phi_2,\tphi)=& V_1(\phi_1) \,  \theta(\tphi) + V_2(\phi_2)\, \theta(-\tphi) \nonumber \\
    &+ \Lambda (\tphi -1)^2(\tphi+1)^2, \\
    C(\phi_1,\phi_2,\tphi) =& B_1(\phi_1)\,  \theta(\tphi)+ B_2(\phi_2)\,  \theta(-\tphi),\\
    \beta(\phi_1,\phi_2,\tphi)=& \alpha_1(\phi_1)\,  \theta(\tphi) + \alpha_2(\phi_2)\,  \theta(-\tphi),
\end{align}
where $\theta$ is the Heaviside function,
we indeed have $\ket{W_{\infty},C,\beta}\propto\ket{V_1,B_1,\alpha_1}+\ket{V_2,B_2,\alpha_2}$.  Indeed, when $\Lambda$ is sent to infinity, the auxiliary field $\tphi$ becomes a ``bit'' taking values $\pm 1$ digitally splitting the functional integral into two contributions $\int \mathcal{D} \tphi \simeq \sum_{\tphi\equiv\pm 1}$ where each term of the sum gives the two initial states.

The expressiveness of cTNS is then easy to assess, following the same technique as for cMPS \cite{jennings2015}. Taking $\alpha(x,\phi(x))=f(x)$ and $V=a/\text{Vol}(\Omega)$ we obtain any field coherent state with any complex weight $\e^{-a}\ket{f}$. Using the stability result, one can construct arbitrary linear combinations of such field coherent states which are dense in Fock space, hence one can get arbitrarily close to any state in the Fock space. With this construction, the bond field-dimension grows at each addition of coherent states. 

Actually, using larger bond field dimensions is only a convenience and, provided $V$ and $\alpha$ are arbitrary, a cTNS can approximate any state in the Fock space with $D=1$. Let us consider a sum of coherent states $\ket{\Psi} = \sum_{j=1}^m \e^{-a_i} \ket{f_i}$. This sum can be approximated by the cTNS $\ket{V_\Lambda,1,\alpha}$ with:
\begin{align}
V_\Lambda(x,\phi(x))=& -\Lambda\,\mathbf{1}_{[-1/2,m+1/2]}[\phi(x)] \cos(2\pi \phi(x))  \nonumber\\
&+\sum_{j=1}^m a_j\mathbf{1}_{ [j-1/2,j+1/2]}[\phi(x)]  \label{eq:V_M}\\
\alpha(x,\phi(x))=&\sum_{j=1}^m\mathbf{1}_{[j-1/2,j+1/2]} [\phi(x)]\,  f_j(x),
\end{align}
where $\mathbf{1}_A[x]=1$ if $x\in A$ and $0$ otherwise.
Indeed, when $\Lambda\rightarrow + \infty$, the auxiliary field is forced to sit on one of the $m$ minima of the potential, which each have a complex weight $\e^{-a_j}$. To each of these $m$ possible values of the field, the term in $\alpha$ associates a different coherent state. Hence one can approximate $\ket{\Psi}$ with arbitrary precision and hence all states in the Fock space.

Allowing larger values of $D$ remains useful if $V$ and $\alpha$ are restricted in some way, \eg to being polynomials with a fixed degree. In that case, being able to take a larger bond field dimensions $D$ substantially increases the expressiveness of a cTNS subclass. Gaussian cTNS (see \ref{sec:gaussianstates}) will provide such an illustration.

\subsection{Gauge transformation}\label{subsec:gaugeinvariance}

\cbl{Different choices of $V$, $B$, and $\alpha$ can generate the same state. This is to be expected: In the discrete, the map between an elementary tensor and a many-body wave-function is not injective either. Understanding the transformations between tensors generating the same state is fundamental in the theory of TNS, especially for the classification of symmetry protected and topological phases. It is thus natural to ask the same question for cTNS following the discrete construction.}

\subsubsection{Intuition from the discrete}
In the discrete, there exists \cbl{an important subclass of} transformations one can apply on the bond indices of an elementary tensor and that leave the state invariant. For example in $d=2$, the transformation (sometimes called \emph{gauge} transformation):
\begin{equation}\label{eq:gauge}
\vcenter{\hbox{\includegraphics[scale=\tsize]{singletensor.pdf}}} \hskip0.4cm \longrightarrow \hskip0.4cm
\vcenter{\hbox{\includegraphics[scale=\tsize]{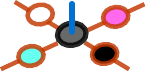}}}
\end{equation}
where $\vcenter{\hbox{\includegraphics[scale=\tsize]{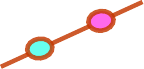}}} = \!\!\!\vcenter{\hbox{\includegraphics[scale=\tsize]{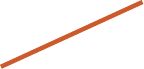}}}$ and $\vcenter{\hbox{\includegraphics[scale=\tsize]{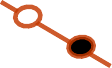}}} =\!\!\! \vcenter{\hbox{\includegraphics[scale=\tsize]{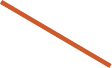}}}$
gives the same contracted state up to new boundary terms (that vanish on a torus). Such transformations have proved central to classify topological phases of matter with discrete tensor networks. We would thus like to find an analog in the continuum.

\cbl{For infinite bond dimension}, the equivalent of an invertible linear transformation acting on discrete indices is a linear operator $\mathsf{G}$ acting on functions of $D$ real variables (the auxiliary field):
\begin{equation}
\mathsf{G}\cdot \varphi (\phi) = \int\upd^D \widetilde{\phi}\; \mathsf{G}(\phi,\widetilde{\phi}) \,\varphi(\widetilde{\phi})
\end{equation}
A $\mathsf{G}$ that is too generic will typically destroy the continuum limit when the corresponding gauge transformation \eqref{eq:gauge} is applied on the elementary tensor. The main difficulty is to know what subset of operators to look at. For simplicity, we restrict ourselves to the case $D=1$ from which the general case is easily deduced.

A first option is to consider the subset of diagonal transformations. Let $\mathsf{F}$ and $\mathsf{G}$ be two operators acting diagonally:
\begin{align}
\mathsf{G}\cdot\varphi(\phi)=\mathsf{g}(\phi)\varphi(\phi)\\
\mathsf{F}\cdot\varphi(\phi)=\mathsf{f}(\phi)\varphi(\phi)
\end{align}
Acting on an elementary discrete tensor in the same way as in \eqref{eq:gauge} with $\mathsf{F}$ and $\mathsf{G}$ simply changes the integration measure:
\begin{equation}\label{eq:measprefactor}
\upd \phi(1)\cdots\upd\phi(4)\rightarrow \frac{f(\phi(1)) g(\phi(4))}{f(\phi(3)) g(\phi(2))}\upd \phi(1)\cdots \upd\phi(4)
\end{equation}
If this change of measure is too general, there will be no continuum limit. A natural choice, preserving the continuum, is to take:
\begin{align}
\mathsf{f}(\phi)&=\exp\left(-\varepsilon^{d-1} \mathfrak{f}(\phi)\right)\\
\mathsf{g}(\phi)&=\exp\left(-\varepsilon^{d-1} \mathfrak{g}(\phi)\right)
\end{align}
In the continuum limit such a choice put in \eqref{eq:measprefactor} yields:
\begin{equation}
\mathcal{D}\phi\rightarrow \mathcal{D}\phi\exp\left[\int \upd^2 x \; \nabla\cdot\left(\begin{array}{c}
\mathfrak{f}(\phi(x))\\
\mathfrak{g}(\phi(x))
\end{array}\right) \right]
\end{equation}
hence this adds a pure divergence term into the cTNS definition which can be transformed into a boundary term thanks to Stokes' theorem. This is exactly what a gauge transformation should do.

This very special choice of operators does not exhaust the infinitesimal transformations compatible with the existence of a continuum limit. However, we conjecture that all discrete gauge transformations of the form \eqref{eq:gauge} that preserve the continuum limit ultimately give rise to pure divergence terms as well. In any case, this discussion of the discrete setting is but a motivation for the introduction of (some) continuous gauge transformations of cTNS.

\begin{figure}
    \centering
    \includegraphics[width=0.99\columnwidth]{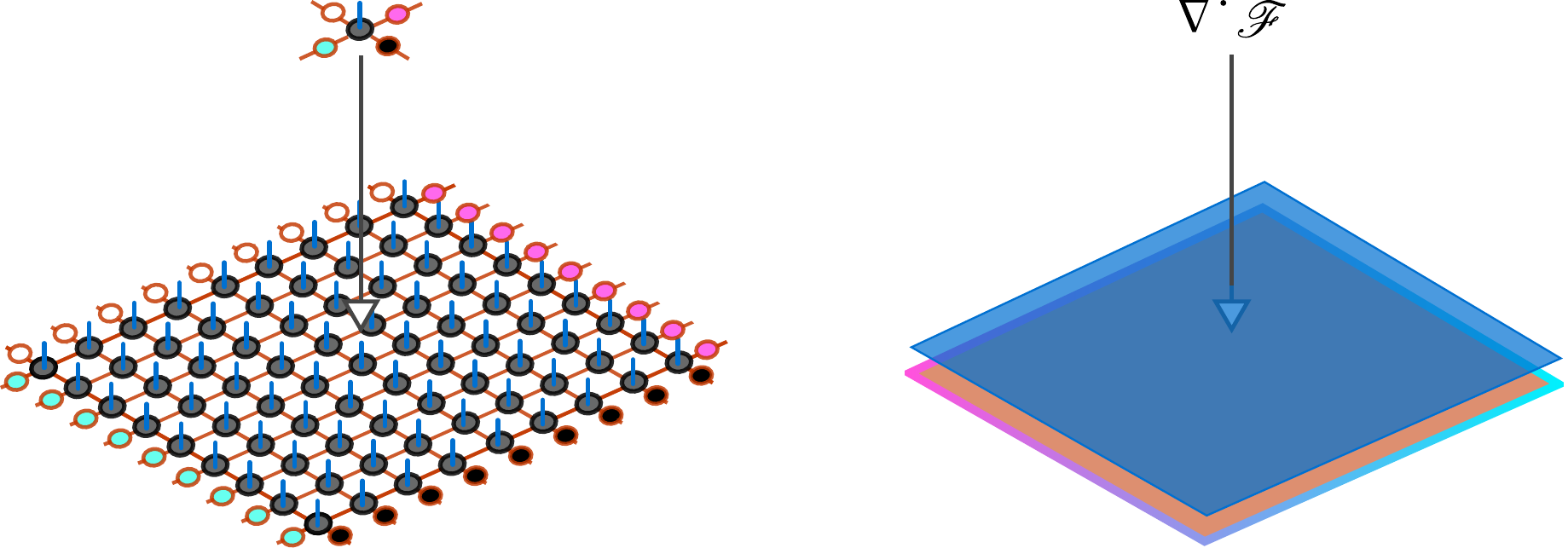}
    \caption{\textbf{Gauge transformations} -- In the discrete case (left), transforming the elementary tensor as in \eqref{eq:gauge} has a non trivial result on the boundary only. In the continuum (right), the transformation of the elementary tensor is equivalent to the addition of a pure divergence term for the auxiliary fields in the bulk, which can then be integrated into a boundary condition.}
    \label{fig:gauge}
\end{figure}

\subsubsection{Continuum description}
The previous inquiries motivate the following proposition which is at the same time a definition of a certain class of gauge transformations for cTNS.
\begin{proposition}[Gauge transformation]
Let $\mathscr{F}[x,\phi(x),\nabla\phi(x)]$ be an arbitrary vector field in $\Omega$. If $\Omega$ has no boundary, the cTNS $\ket{V,\alpha}$ is left unchanged by the gauge transformation:
\begin{equation}\label{eq:gaugetransform}
V(\phi)\rightarrow V(\phi) + \nabla\cdot \mathscr{F}[x,\phi(x)].
\end{equation}
\end{proposition}

The proof is trivial and is just a direct application of Stokes' theorem.
More generally, if $\Omega$ has a boundary $\partial\Omega$, the gauge transformation \eqref{eq:gaugetransform} adds a boundary term to the measure:
\begin{equation}
B(\phi) \rightarrow B(\phi) \exp\left\{ \oint_{\partial\Omega} \!\!\upd^{d-1}x\, \mathscr{F}[x,\phi(x)] \cdot \mathbf{n}(x)\right\}.
\end{equation}
where $\mathbf{n}(x)$ is the unit vector normal to $\partial\Omega$ in $x$. Gauge transformations of cTNS thus have a straightforward geometric interpretation.

\subsection{Tensor rescaling}\label{subsec:renormalization}
Our objective is now to relate different tensor network descriptions of the same state at different scales \cite{verstraete2004-renormalization,vidal2007}. More precisely, considering a correlation function for a state parameterized by a tensor $T(1)$ in the thermodynamic limit:
\begin{equation}
C(x_1,\cdots,x_n) = \bra{T(1)} \mathcal{O}(x_1)\cdots \mathcal{O}(x_n) \ket{T(1)},
\end{equation}
the objective is to find a tensor $T(\lambda)$ of new parameters such that:
\begin{equation}
C(\lambda x_1,\cdots,\lambda x_n) \propto \bra{T(\lambda)} \mathcal{O}(x_1)\cdots \mathcal{O}(x_n)\ket{T(\lambda)}.
\end{equation}
Naturally, in the discrete, this relation is at best approximate.

For the cTNS of definition \ref{def:pathintegral}, we can write the flow $V(\lambda), \alpha(\lambda)$ exactly, following the rather standard dimensional analysis of ordinary QFT. As such, we introduce new creation and annihilation operators $\widetilde{\psi}^\dagger(x)=\lambda^{d/2} \psi^\dagger(x\lambda)$ and $\widetilde{\psi}(x)=\lambda^{d/2} \psi(x\lambda)$. They indeed verify the standard commutation relations $[\widetilde{\psi}(x),\widetilde{\psi}^\dagger(x)]=\delta(x-y)$. These new operators relate correlation functions at different scales:
\begin{align}
C(\lambda x_1,\cdots, \lambda x_n) :&=\langle \psi^\dagger(\lambda x_1) \cdots \psi(\lambda x_n)\rangle\\
&= \lambda^{nd/2} \langle \widetilde{\psi}^\dagger(x_1)\cdots \widetilde{\psi}^\dagger(x_n)\rangle,
\end{align}
where the $\lambda^{nd/2}$ factor just comes from the fact that the operators we introduced have a dimension.
We just have to rewrite $\ket{V,\alpha}$ as a function of the new creation and annihilation operators to relate the different scales. To achieve this, we change of position variable introducing $u=x/\lambda$. The free field measure now reads:
\begin{align}
\upd \mu(\phi)&=\mathcal{D}\phi\exp\left(-\frac{1}{2}\int\upd^d u \,\lambda^{d-2} \;\nabla \phi_k (u\lambda) \cdot \nabla \phi_k (\lambda u)\right)\\
&\propto \mathcal{D}\widetilde{\phi}\exp\left(-\frac{1}{2}\int\upd^d u \, \;\nabla \widetilde{\phi}_k (u) \cdot \nabla \widetilde{\phi}_k (u)\right) \\
&= \upd \mu(\widetilde{\phi}),
\end{align}
with $\widetilde{\phi}(u) =\lambda^{\frac{d-2}{2}} \phi(\lambda u)$. 
This gives:
\begin{align}
    \ket{V,\alpha}_\psi &=
\int \upd \mu(\widetilde{\phi}) \exp\Big\{-\int \upd^d u \; \lambda^d V[\lambda^{\frac{2-d}{2}}\widetilde{\phi}(u)] \nonumber\\
    &\hskip1.5cm- \lambda^{\frac{d}{2}}\alpha[\lambda^{\frac{2-d}{2}}\widetilde{\phi}(u)] \, \widetilde{\psi}^\dagger(u)\Big\} \; \ket{0} \\
    &=\ket{\lambda^d V[\lambda^{\frac{2-d}{2}}\;\cdot\;],\lambda^{\frac{d}{2}}\alpha[\lambda^{\frac{2-d}{2}}\;\cdot\; ]}_{\widetilde{\psi}}.
\end{align}
This allows to discuss the IR behavior of cTNS in terms of \emph{relevant}, \emph{irrelevant}, and \emph{marginal} couplings. To this end, we informally expand $V$ and $\alpha$ in powers $p$ of the fields $\phi$ and analyze the terms of each degree separately. The corresponding coupling dimensionality is $\Delta= d + \frac{2-d}{2} p$ for terms in $V$ and $\Delta=\frac{d}{2} +\frac{2-d}{2} p$ for $\alpha$. Consequently:
\begin{itemize}
\item[--] For $d=2$, All powers of the field in $V$ and $\alpha$ yield relevant couplings. There are no irrelevant couplings.
\item[--] For $d=3$, the powers $p=1,2,3,4,5$ of the field in $V$ yield relevant $\Delta>0$ couplings. The power $p=6$ is marginal in $V$. For $\alpha$, the powers $p=1,2$ are relevant and $p=3$ is marginal. All other powers are irrelevant.
\end{itemize}
Relevant powers will dominate the behavior of cTNS correlation functions in the IR and actually be the only ones allowed if the cTNS description is aimed to hold at all scales in the non Gaussian case (see \ref{subsec:nongaussian}).

\subsection{Recovering continuous matrix product states}\label{subsec:recovering_cmps}
Continuous TNS should reduce to cMPS in an appropriate limit. This is an important property to check to demonstrate that our ansatz is a natural extension of cMPS to $d\geq 2$.

\subsubsection{Compactification}

A cMPS of a quantum field defined on a space interval of length $T$ is parametrized by $3$ $(\chi\times \chi)$ matrices $\hat{Q},\hat{B},\hat{R}$ and is defined \cite{verstraete2010,haegeman2013} as:
\begin{equation}
    \ket{Q,B,R} = \tr\left\{\hat{B}\mathcal{T}\exp\left[\int_{-T/2}^{T/2}\!\!\!\upd\tau\, \hat{Q} + \hat{R}\otimes \psi^\dagger(\tau)\right]\right\}\ket{0}.
\end{equation}

To obtain a cMPS, we may directly instantiate our cTNS ansatz with $d=1$, \eg using its functional integral form \eqref{eq:pathintegral}. However, as we mentioned before, the $d=1$ case is quite peculiar compared to other dimensions and so it is nice to see it can also be immediately obtained from a general $d$ case where all dimensions but one are taken to be very small.

Indeed, consider a domain of the form $\Omega = [-T/2,T/2]\times S$ where $S$ is a $d-1$ dimensional torus of length $\ell$ in all $d-1$ directions. Expanding the auxiliary fields $\phi$ of \eqref{eq:pathintegral} in Fourier modes on this $d-1$ torus $S$ and taking the limit $\ell\rightarrow 0$, yields a functional integral in which only the field zero mode on $S$ survives. Hence, one obtains a functional integral of the form: 
\begin{equation}
\begin{split}
    \ket{V,B,\alpha} = \int \mathcal{D} \phi B(\phi)& \exp\bigg\{-\!\int_{-T/2}^{T/2}\!\!\!\upd x \; \frac{1}{2} \sum_{k=1}^D [\partial_x \phi_k(x)]^2  \\
    &+ V[\phi(x)] - \alpha[\phi(x)] \, \psi^\dagger(x)\bigg\} \ket{0},
\end{split}
\end{equation}
where the $1$ dimensional auxiliary field $\phi$ is the zero mode (on the shrunk torus $S$) of the initial $d$ dimensional auxiliary field. It is what one would have obtained fixing immediately $d=1$ in \eqref{eq:pathintegral}. In operator form, this yields:
\begin{equation}\label{eq:particle}
\begin{split}
\ket{V,B,\alpha}=\tr \bigg[\hat{B} \mathcal{T} \exp\Big(&\!\!-\!\!\int_{-T/2}^{T/2} \!\!\!\upd \tau  \sum_{k=1}^D\frac{\hat{P}_k^2}{2} \\
+& V[\hat{X}] - \alpha[\hat{X}]\,\psi^\dagger(\tau) \Big)\bigg]\ket{0}
\end{split}
\end{equation}
where $\hat{P}_k$ and $\hat{X_k}$ are canonically conjugated pairs ($D$ zero dimensional quantum fields). This is already a cMPS with $-\hat{Q} = \sum_{k=1}^D\frac{\hat{P}_k^2}{2}
+ V[\hat{X}]$ and $\hat{R}=\alpha[\hat{X}]$. However, the bond Hilbert space is now that of $D$ particles in $1$ dimension or $1$ particle in $D$ dimensions, hence $\chi = +\infty$. 

\subsubsection{Bond dimension quantization}

To obtain a genuine cMPS (with finite bond dimension) from a $d=1$ cTNS defined by \eqref{eq:particle}, we need to choose a specific potential effectively reducing the Hilbert space dimensionality. The intuition is quite clear: take a potential with deep minima.

Let us take a potential with $D$ deep minima $m_k$ on the vertices closest to $0$ of an hypercube, \ie $m_1=(1,0,\cdots,0)$, $m_2=(0,1,0,\cdots,0)$, ..., and $m_D=(0,\cdots,0,1)$. The effective dynamics is now restricted to a $D$-dimensional Hilbert space spanned by $\ket{m_k}$ corresponding to wave packets localized around each minima. In this reduced Hilbert space, the minima are coupled by tunneling. Because of the geometrical configuration we have considered, the minima can all be connected by independent saddle points, hence the effective coupling between the minima can be chosen freely. This means that we can obtain any $D\times D$ complex matrix $Q$ of standard cMPS by adjusting the value of the $D^2$ saddle points of $V$. 

The $R$ matrix is fixed in the same way. The values of $\alpha[X]$ on the minima $\ket{m_k}$ of the potential fix the diagonal coefficients of $R$ and the value on the saddle point connecting $\ket{m_{k_1}}$ and $\ket{m_{k_2}}$ fix the non diagonal terms $R_{k_1,k_2}$.

Hence, not only can cTNS reduce to cMPS when $d=1$ (or when $d-1$ dimensions are small) for a specific choice of potential, but actually all (bosonic) cMPS can be obtained this way. In this context, the bond field dimension $D$ reduces to the usual bond dimension $\chi$. 

\section{Computations}\label{sec:computations}
\cbl{To carry computations with cTNS, one could of course rediscretize them and use the standard TNS algorithms. We now mention techniques relying only on the continuum limit.}

\subsection{Gaussian states} \label{sec:gaussianstates}
There exists a subclass of cTNS for which all quantities of interest can be computed exactly: Gaussian cTNS.
\begin{definition}[Gaussian cTNS]\label{def:gaussian}
A cTNS is said to be \emph{Gaussian} if the functions $V$ and $\alpha$ are respectively at most quadratic and affine in the auxiliary field:
\begin{align}
V(x,\phi)&=V^{(0)}(x) + V^{(1)}_k(x)\,\phi_k+ \frac{1}{2}V^{(2)}_{k\ell}(x)\, \phi_k \phi_\ell \\
\alpha(x,\phi)&= \alpha^{(0)}(x) + \alpha^{(1)}_k(x)\,\phi_k
\end{align}
\end{definition}
Naturally, a Gaussian cTNS is also a Gaussian state in the usual sense of the term. More precisely, for a Gaussian cTNS, $\mathcal{Z}_{j',j}$ and $\widetilde{\mathcal{Z}}_{j',j}$ are manifestly Gaussian functionals. Let us compute $\mathcal{Z}_{j',j}$ in the translation invariant case $\Omega=\mathds{R}^d$. 
\begin{widetext}
Inserting definition \ref{def:gaussian} into equation \eqref{eq:ordered} yields:
\begin{align}
\mathcal{Z}_{j',j}=\frac{1}{\mathcal{N}} \int \mathcal{D}[\phi]\mathcal{D}[\phi'] &\exp\bigg\{ -\int \upd^{d}x \left(\begin{array}{c}
\phi\\
\phi'
\end{array}\right)^T \!\!\!\cdot 
\left(\begin{array}{cc}
\frac{-\triangle+V^{(2)}}{2} & \frac{-\alpha^{(1)}\otimes \alpha^{(1)*}}{2}\\
\frac{-\alpha^{(1)*}\otimes\alpha^{(1)}}{2}  & \frac{-\triangle+V^{(2)*}}{2} 
\end{array}\right) \cdot\left(\begin{array}{c}
\phi\\
\phi'
\end{array}\right)\nonumber \\
& + \left(\begin{array}{c}
V^{(1)} - \alpha^{(1)}(j+\alpha^{(0)*}) \\
V^{(1)} - \alpha^{(1)*}(j'+ \alpha^{(0)})
\end{array}\right)^T \!\!\!\cdot
\left(\begin{array}{c}
\phi\\
\phi'
\end{array}\right) - j\alpha^{(0)} - j' \alpha^{(0)*} - \alpha^{(0)} \alpha^{(0)*}+ V^{(0)} + V^{(0)*}\bigg\}.
\end{align}
Carrying the Gaussian integration we then obtain:
\begin{equation}
\begin{split}
\mathcal{Z}_{j,j'}=\frac{1}{\widetilde{\mathcal{N}}}\exp\bigg\{
\int \upd^d x\,\upd^d y \;\frac{1}{2} \; \Lambda(j,j')^T(x) \cdot K(x,y) \cdot \Lambda(j,j')(y) + 
\delta(x-y)(j\alpha^{(0)} + j' \alpha^{(0)*})(y)\bigg\}
\end{split}
\end{equation}
where:
\begin{align}
 &\Lambda(j,j')=\left(\begin{array}{c}
V^{(1)} - \alpha^{(1)}(j+\alpha^{(0)*}) \\
V^{(1)} - \alpha^{(1)*}(j'+ \alpha^{(0)})
\end{array}\right)\text{ and } \left(\begin{array}{cc}
-\triangle+V^{(2)} & -\alpha^{(1)}\otimes \alpha^{(1)*}\\
-\alpha^{(1)*}\otimes\alpha^{(1)}  & -\triangle+V^{(2)*}
\end{array}\right)K(x,y) = \mathds{1}_{2D\times 2D} \delta(x-y). \label{eq:kernel}
\end{align}
Because of translation invariance, $K(x,y)=K(x-y)$ which can be written in Fourier space:
\begin{equation}
K(x-y)=\int \upd^{d} p \;\e^{ip\cdot (x-y)} K(p).
\end{equation}
Inserting this expression into equation \eqref{eq:kernel} and integrating over the variable $u=(x-y)$ yields:
\begin{equation}
K(p)=\frac{1}{(2\pi)^d} \left(\begin{array}{cc}
p^2+V^{(2)} & -\alpha^{(1)}\otimes \alpha^{(1)*}\\
-\alpha^{(1)*}\otimes\alpha^{(1)}  & p^2+V^{(2)*}
\end{array}\right)^{-1},
\end{equation}
which is difficult to make more explicit but could be computed exactly for given $\alpha$ and $V$. To get a intuition of the behavior of the two point functions, we may instantiate this expression on a simple example where the bond field-dimension $D$ equals $1$ and $\alpha^{(0)}=V^{(1)}=0$ for simplicity. In that case we have:
\begin{equation}
K(p)=\frac{1}{(2\pi)^d} \frac{1}{(p^2+V^{(2)})(p^2+V^{(2)*})- |\alpha^{(1)}|^4} \left(\begin{array}{cc}
p^2+V^{(2)*} & |\alpha^{(1)}|^2\\
|\alpha^{(1)}|^2  & p^2+V^{(2)}
\end{array}\right).
\end{equation}
Using \eqref{eq:correlexample} this gives the correlation function:
\begin{align}
C(x-y):=\langle \psi^\dagger(x) \psi(y)\rangle&=\frac{|\alpha^{(1)}|^2}{2}\left[ (1,0) K(x,y)(0,1)^T + (0,1) K(x,y) (1,0)^T\right]\\
&=\frac{1}{(2\pi)^d}\int \upd^d p\, \frac{|\alpha^{(1)}|^4  \,\e^{ip\cdot (x-y)}}{(p^2+V^{(2)})(p^2+V^{(2)*})- |\alpha^{(1)}|^4}.
\end{align}
Importantly here, the correlation function in momentum space $C(p) \propto p^{-4}$ when $p\rightarrow +\infty$. Hence, the integral is not UV divergent for $x=y$ so long as $d\leq 3$ in which case the particle density $\langle \psi^\dagger(x) \psi(x)\rangle$ is finite. 
\end{widetext}

\subsection{Non-Gaussian states} \label{subsec:nongaussian}
For a non-Gaussian cTNS, it is no longer possible to compute the correlation functions exactly in general. Further, the definitions we provided for the cTNS in \eqref{eq:pathintegral} or \eqref{eq:operator} are generically divergent. Nonetheless one can use approximations or numerical techniques coming from the quantum field theory and tensor network toolboxes.

\subsubsection{Regularization and renormalization}

In the general case, the ansatz we put forward suffers from the same UV divergences that plague quantum field theories. As in QFT, these divergences are in a way inevitable: the gradient squared $(\nabla\phi)^2$ in the path integral  insufficiently penalizes high momenta in $d\geq 1$ (note that, again, the $d=1$ case is trivial). On the other hand, the locality of the underlying tensor network forbids higher derivatives. Hence, as in QFT, the divergences are tied to the very property (locality) that we require.

Given this state of affairs, there are essentially $3$ options to deal with divergences, depending on what one needs the state for.

The first option is simply to regularize the state with a momentum cutoff $\Lambda$, either directly in the path integral --which will break locality and destroy the operator representation-- or in the operator representation --which will generically break Euclidean invariance--. In both cases, the scale $\Lambda$ will be reminiscent of the inverse lattice spacing of discrete tensor networks. The parameters appearing in the expansion of $V$ and $\alpha$ will then be the equivalent of the bare parameters in QFT Lagrangians. As long as the state is used as a variational ansatz \eg to minimize the energy of an anyway regularized QFT Hamiltonian, this is unproblematic. Indeed, carrying an optimization on bare or renormalized parameters will be equivalent, and the fact that some properties break above a cutoff momentum is anyhow imposed by the physical QFT being approximated. In this approach, there is no restriction on the powers of the auxiliary field appearing in $V$ and $\alpha$.

One may also be interested in the class of cTNS for their properties, and not necessarily to approximate the ground state of a given system. In that case, going beyond regularization and renormalizing the state with proper counterterms and renormalization conditions seems necessary to preserve the locality of the underlying tensor network. In the general case, this is equivalent to renormalizing a relativistic open quantum field theory, a problem which has received interest recently \cite{baidya2017}. At the level of dimensional analysis, this restricts powers of the auxiliary field in $V$ and $\alpha$ to renormalizable interactions, hence to the relevant and marginally relevant powers obtained in \ref{subsec:renormalization}. In $d=3$, this restricts the parameters to a finite number of tensors appearing in the finite polynomial expansion of $V$ and $\alpha$. Allowing for more auxiliary fields is thus necessary to make the cTNS class arbitrarily large and expressive in $d=3$. 

Finally, a natural regularization may be provided by restricting the class of quantum states in $d-1$ on which the transfer matrix $\mathds{T}$ acts. As we will see, for special cases of $\mathds{T}$, one can indeed recover finite results in $d=2$. In that case, the state itself is implicitly defined by the approximate method used to contract it.

\subsubsection{Dimensional reduction}
We now discuss the last option. To compute physical correlation functions in the general case, one can exploit their operator expression \eqref{eq:operator-ordered} given by the exponential of a transfer matrix acting in a space of one dimension less. In the $d=2$ case, the theory one needs to solve is thus simply a $1$ dimensional QFT. The latter is solvable with cMPS (\ie cTNS in $1$ dimension less) which, as a bonus, have a built-in UV regulator~\cite{haegeman2010}  and bring the computation back to a $0$ dimensional problem~\cite{osborne2010}. We outline the steps of such a computation on a simple example.

We consider a cTNS on a torus $(\tau,x)$, $\tau \in [0,T]$, $x\in [0,L]$ (as in Fig. \ref{fig:operator}), which reads, in the operator representation:
\begin{equation}
\begin{split}
\ket{V,\alpha}=\tr \bigg[\mathcal{T} \exp\Big(-&\int_{0}^{T} \!\!\!\upd \tau \!\int_0^L\! \!\upd x \; \mathcal{H}(x)\\
-& \alpha[\hat{\phi}(x)]\,\psi^\dagger(\tau,x) \Big)\bigg]\ket{0}.
\end{split}
\end{equation}
Correlation functions for this state at a fixed $\tau$ take a particularly simple form, with the propagator $\exp(\tau \mathds{T})$ appearing only once. For example, the 2-point function reads:
\begin{equation}
\begin{split}
    \langle \psi^\dagger(\tau,x_1) \psi(\tau,x_2)\rangle = \tr \left[\mathds{1} \otimes \hat{\alpha}^*(x_1) \cdot \hat{\alpha}(x_2)\otimes \mathds{1} \cdot \e^{T \mathds{T}} \right].
\end{split}
\end{equation}
Such $N$-point functions at equal $\tau$ contain useful information about the state in the thermodynamic limit $T \rightarrow +\infty$, $L\rightarrow + \infty$. Indeed, because of Euclidean invariance, they give access to all correlation functions of aligned points, and a fortiori to all possible $2$-point functions. This is sufficient to compute the expectation values of most homogeneous and isotropic quasi-local Hamiltonians. 

To simplify the discussion, we now consider the special case of an Hermitian $\mathds{T}$ (obtained \eg when all the coefficients of $V$ and $\alpha$ are real). In the $T\rightarrow +\infty$ limit, $\exp(T \mathds{T})$ will be dominated by the projector on the eigenvector $\ket{\mathrm{ss}}$ of $\mathds{T}$ with the largest eigenvalue. The correlation functions then simplify, \eg:
\begin{equation}\label{eq:physical_stationary}
\begin{split}
    \langle \psi^\dagger(\tau,x_1) \psi(\tau,x_2)\rangle\propto \bra{\mathrm{ss}}\mathds{1} \otimes \hat{\alpha}^*(x_1) \cdot \hat{\alpha}(x_2)\otimes \mathds{1} \ket{\mathrm{ss}}.
\end{split}
\end{equation}
The right hand side is the correlation function for ($2D$ copies of) a $1$ dimensional bosonic field theory, which motivates the use of a cMPS. More precisely, we may use a cMPS defined on two copies of the auxiliary quantum fields to approximate the dominant eigenvector $\ket{\mathrm{ss}}$. Assuming only $D=1$ auxiliary field, we can write:
\begin{equation}
\begin{split}
    \ket{{Q,R_1,R_2}}=&\tr\bigg[\mathcal{P}_x\exp \Big\{\int_0^L \!\!\upd x \, Q\otimes \mathds{1} \\
    &+ R_1 \otimes \psi^\dagger_1(x) + R_2 \otimes \psi^\dagger_2(x)\Big\} \bigg] \ket{0}
\end{split}
\end{equation}
where $Q$, $R_1$, and $R_2$ are $(\chi \times \chi)$ matrices and $\psi^\dagger_1=\psi^\dagger_{[d-1]}\otimes \mathds{1}$ and $\psi^\dagger_2=\mathds{1}\otimes \psi^\dagger_{[d-1]}$ are the creation operators associated to each copy of the Fock space on which $\mathds{T}$ acts, $\ket{0}$ is the Fock vacuum of these two copies, and the trace is taken over the matrices. This is nothing but a translation invariant cMPS for two species of bosons. The dominant eigenvector can then be approximated by choosing
\begin{equation}\label{eq:argmax}
    Q, R_1, R_2 = \argmax_{Q,R_1,R_2} \frac{\bra{Q,R_1,R_2}\mathds{T}\ket{Q,R_1,R_2}}{\langle Q,R_1,R_2|Q,R_1,R_2\rangle}.
\end{equation}
The right hand side of \eqref{eq:argmax} can be computed explicitly as a function of $Q,R_1,R_2$. Indeed, in the same way as we computed the normal ordered correlation functions for a cTNS in \eqref{eq:operator-ordered}, one can compute the normal ordered correlation functions for a cMPS, replacing $-\mathcal{H}(x)$ by $Q$ and $\alpha(x)$ by $R_1, R_2$ \cite{haegeman2013}, \eg for $x\geq y$:
\begin{equation}
    \langle \psi^\dagger_1 (x) \psi_1(y) \rangle= \tr \left[\e^{(L-x)\trans} (\mathds{1}\otimes R_1^*)e^{(x-y)\trans} (R_1\otimes\mathds{1}) \e^{y \trans}\right]
\end{equation}
with the ($0$-dimensional) transfer matrix:
\begin{equation}
    \trans = Q\otimes \mathds{1} + \mathds{1}\otimes Q^* + R_1 \otimes R_1^* + R_2\otimes R_2^*. 
\end{equation}
One then just has to express $\mathds{T}$ as a function of $\psi_1$ and $\psi_2$ instead of the field and conjugate momenta, which requires a choice, \eg:
\begin{align}
 \hat{\phi}(x)&=\Lambda_0^{-1/2}\frac{\psi_{[d-1]}(x)+\psi_{[d-1]}^\dagger(x)}{\sqrt{2}},\\
    \hat{\pi}(x) &=\Lambda_0^{1/2}\;\, \frac{\psi_{[d-1]}(x)-\psi_{[d-1]}^\dagger(x)}{\sqrt{2}i},
\end{align}
for some $\Lambda_0$.
Taking the expectation value of products of local operators on the cMPS yields divergent contributions. They can be removed \eg by normal ordering $\mathcal{H}$ and $\alpha$ \footnote{While normal ordering $\mathcal{H}$ just amounts to a change of normalization in the functional integral, normal ordering $\alpha$ requires explicit counter terms.} in the operator representation of \eqref{eq:operator}, or by adding a counter term in the Hamiltonian as in \cite{stojevic2015}. In the end, the expectation value to maximize can be written:
\begin{equation}
    \frac{\bra{Q,R_1,R_2}\mathds{T}\ket{Q,R_1,R_2}}{\langle Q,R_1,R_2|Q,R_1,R_2\rangle} = \frac{\tr\left[M(Q,R_1,R_2) \e^{L \trans}\right]}{\tr\left[ \e^{L \trans}\right]},
\end{equation}
where $M(Q,R_1,R_2)$ is some polynomial of $Q$, $R_1$, and $R_2$ explicitly calculable from $V$ and $\alpha$. This expression can be simplified in the thermodynamic limit and then be maximized \eg by gradient ascent \cite{haegeman2013}. In practice, for transfer matrices with relativistic $\mathcal{H}$ like the ones we consider, this maximization has to be carried over matrices $Q,R_1,R_2$ with a fixed maximum norm (or with a soft penalization of large norms). This is necessary to prevent the cMPS and its finite entanglement from capturing only the UV features of the stationary state \cite{haegeman2010}. For sufficiently large bond dimension $\chi$, and taking into account this subtlety, we expect to get a good estimate of the stationary state. Once $Q$, $R_1$, and $R_2$ are fixed this way, physical correlation functions can be computed analytically using equation~\eqref{eq:physical_stationary}. For example, if $\alpha$ is linear $\alpha(\phi) \propto \phi$ we get, for $x\geq y$:
\begin{equation}
\begin{split}
    \langle \psi^\dagger(\tau,x) \psi(\tau,y)\rangle\propto \tr& \Big[\e^{(L-x+y)\trans} (R_2\otimes \mathds{1} + \mathds{1} \otimes R_2^*)\\
    &\times \e^{(x-y)\trans} (R_1\otimes\mathds{1} + \mathds{1}\otimes R_1^*) \Big].
    \end{split}
\end{equation}
More complicated cases could be treated in a similar way. Through $2$ successive dimensional reduction, we can thus compute certain correlation functions of a cTNS in $d=2$ with an expression involving only matrices with a finite number of entries ($d=0$). There is a priori no objection in principle to contract a  $d=3$ cTNS this way, but each additional dimensional reduction is done at the price of a variational optimization.
For numerical purposes, the optimization of the cMPS is the crucial step. While current methods \cite{haegeman2010,stojevic2015,draxler2017,ganahl2018} can be used, the prospect to use cMPS to solve field theories in more than 1 spatial dimension provides a strong additional motivation to make them more efficient.

\subsubsection{Perturbation theory}
Given that it is possible to compute correlation functions for Gaussian cTNS, it is natural to compute correlation functions for more general states by carrying a perturbative expansion around Gaussian states. One simply Dyson expands the non-Gaussian part of the exponential in the expression for the generating functional \eqref{eq:ordered}. It generically yields an expansion in terms of Feynman diagrams, similar to that of QFT. 

For example, if $\alpha[\phi]$ is linear in $\phi$ with a correction $\propto \lambda_{k\ell} \phi_k \phi_\ell$, the expansion will contain diagrams composed of vertices with $3$ and $4$ legs, corresponding to the $\alpha[\phi] \alpha^*[\phi']$ term in \eqref{eq:ordered}, connected by Gaussian propagators. As previously mentioned, unless cancellations between different auxiliary fields occur, loop diagrams will be UV divergent and a regularization will be needed. We leave the derivation of the general Feynman rules, including a renormalization scheme, to future work. Notice that, in this approach, it is not the state itself that is defined through a perturbative expansion, but rather the correlation functions computed with it.

\subsubsection{Others}
There are of course many other ways one could compute correlation functions. As we mentioned before, one could rediscretize the cTNS to go back to a tensor network description, truncate the bond dimension, and use existing algorithms to contract it. However this would seem to partially defeat the purpose of introducing the continuum in the first place. An interesting avenue is to explore known approximations or tools of quantum field theory (besides perturbation theory) that would not be obvious in the discrete, like saddle point approximations, large $D$ limits, or functional renormalization. Finally, direct Monte-Carlo sampling of the auxiliary field, although it will yield oscillating terms harming convergence in the general case, is a last resort option.

\section{Generalizations}\label{sec:generalizations}

\subsection{General metric and anisotropy}

The main difficulty to overcome in order to construct cTNS lay in preserving local Euclidean symmetries. We may now wish to relax this constraint by allowing a general metric and anisotropic terms in the functional integral definition \eqref{eq:pathintegral}. Namely, it is natural to consider the following generalization.
\begin{definition}[General functional integral formulation]\label{def:generalpathintegral}
A continuous tensor network state (cTNS) of a bosonic quantum field on a smooth Riemanian manifold $\mathcal{M}$ with boundary $\partial \mathcal{M}$ and metric $g$, is a state $\ket{V,B,\alpha}$ parameterized by 2 functions $V$ and $\alpha$: $\mathds{R}^{D+dD}\rightarrow \mathds{C}$, and a boundary functional $B$: $L^2(\partial M) \rightarrow \mathds{C}$ defined by the functional integral on an auxiliary $D$-component field $\phi$:
\begin{equation}\label{eq:generalpathintegral}
\begin{split}
    &\ket{V,B,\alpha} = \int \mathcal{D} \phi \, B(\phi|_{\partial \mathcal{M}})  \exp\bigg\{\!\!-\!\!\int_\mathcal{M} \upd^d x \sqrt{g} \\
    &\times\Big(\frac{g^{\mu\nu}\partial_\mu \phi_k\partial_\nu \phi_k}{2} 
    + V[\phi,\nabla\phi]- \alpha[\phi,\nabla\phi] \, \psi^\dagger\Big)\bigg\} \; \ket{0},
\end{split}
\end{equation}
where all functions depend explicitly on position and summation on 
$k$ is assumed.
\end{definition}

\subsection{Specialization: cMERA} \label{subsec:cmera}
\begin{figure}
    \centering
    \includegraphics[width=0.99\columnwidth]{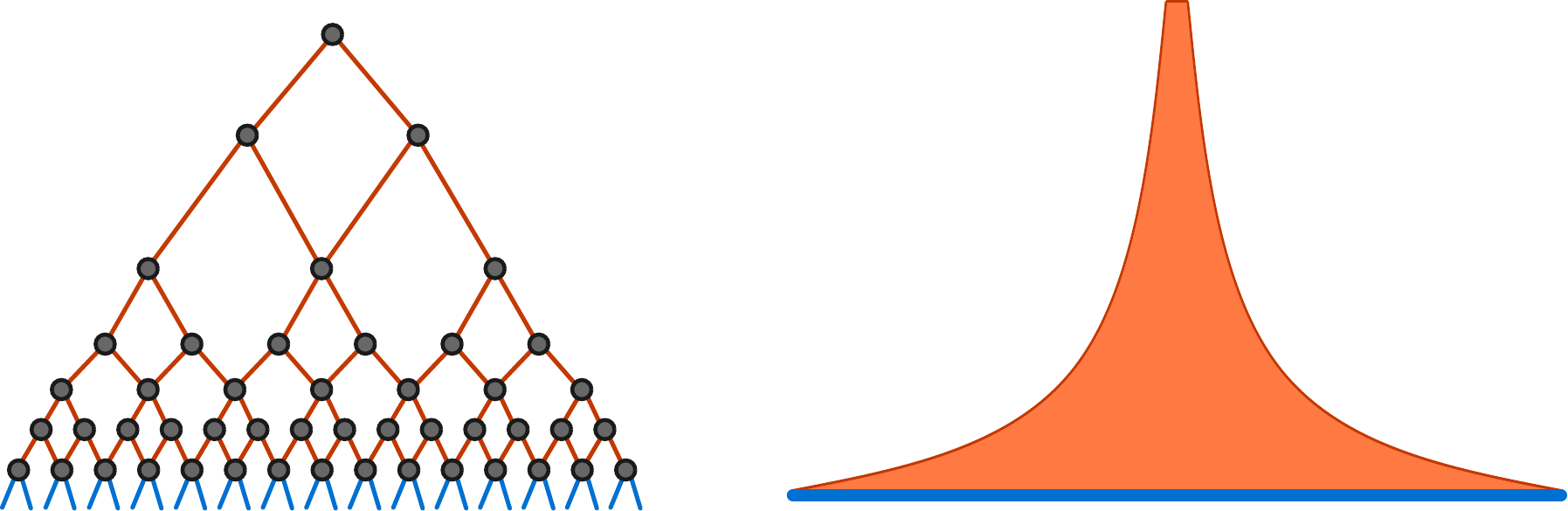}
    \caption{\textbf{Physical states on a boundary} -- In the discrete (left), the MERA is a (special case of) tensor network state with a hierarchical structure, with physical indices only at the boundary.  Tensor network states with such a structure can also be extended to the continuum with our cTNS ansatz, by restricting the physical field to the boundary and choosing an appropriate metric (here hyperbolic) for the bulk auxiliary fields.}
    \label{fig:cmera}
\end{figure}

A natural specialization of the previous generalization consists in having an auxiliary field living on an hyperbolic manifold $\mathcal{M}$ coupled to a physical field restricted to the boundary $\partial \mathcal{M}$ (see Fig. \ref{fig:cmera}). Namely, we have in mind a state of the form:
\begin{equation}\label{eq:cmera}
\begin{split}
    \ket{V,\alpha}\sim\!\int\! \mathcal{D}\phi\, &\exp\bigg\{\!\!-\!\!\int_\mathcal{M}\!\!\! \sqrt{g} 
    \Big(\frac{g^{\mu\nu}\partial_\mu \phi_k\partial_\nu \phi_k}{2} 
    + V[\phi]\Big)\bigg\} \\ \times &\exp\bigg\{\oint_{\partial\mathcal{M}}\!\!\! \alpha[\phi]\,\psi^\dagger\bigg\}\; \ket{0}
\end{split}
\end{equation}
where the integral on the boundary may have to be taken as some appropriately rescaled limit of a bulk integral.
Such states could provide a natural generalization of the multi-scale entanglement renormalization ansatz (MERA) \cite{vidal2008} in the continuum and for an arbitrary number of physical dimensions. They could provide a natural continuum versions of tensor network toy models of the AdS/CFT correspondence \cite{swingle2012,pastawski2015,hayden2016}. The form \eqref{eq:cmera} is also reminiscent of field theory toy models of the AdS/CFT correspondence \cite{penedones2015}, where scalar field theories on a fixed AdS background are related to conformal field theories on the boundary.

Note that this is approach is different in spirit from that of the standard entanglement renormalization approach to quantum fields \cite{haegeman2013-cmera, nozaki2012,mollabashi2014,cotler2018-short,cotler2018-long}, constructed as a unitary transformation applied on a QFT ground state. In the proposal \eqref{eq:cmera}, there is a straightforward lattice discretization and a natural ``bulk'' description in terms of auxiliary fields. However, the isometry property, characteristic of the MERA, is less straightforward to implement.

\subsection{Fermions}
We have defined our ansatz for bosonic quantum fields, because functional integrals and field coherent states are more natural in this context. To extend our proposal to fermions, one would have to introduce quite peculiar Grassmanian integrals with even kinetic and potential $V$ terms but a Grassman odd term $\alpha$ in front of the creation operator $\psi^\dagger$. For fermions, it may be more convenient to start with an operator representation like that of \eqref{eq:operator}, where Euclidean invariance is less natural, to subsequently derive the functional integral formulation.

\subsection{Conformal field theory}
We defined cTNS with the help of $D$ auxiliary free massless scalar fields of measure $\upd \mu$. A natural generalization would be to consider more general conformal field theories (CFT) for the auxiliary space, in the spirit of what has been proposed in the context of matrix product states with infinite bond dimensions \cite{cirac2009}. Admittedly, some non-trivial measures can already effectively be emulated by tuning the real part of the potential $V$ in \eqref{eq:pathintegral}. However in the general case, it may be more convenient to use the CFT machinery directly, for example on the wave function representation (\ref{eq:wave-path},\ref{eq:wave-operator}) or on correlation functions.

\section{Discussion}
We have put forward a new class of states for quantum fields that is obtained as a continuum limit of tensor network states and thus carries the same fundamental properties. 

Although we have shown a number of interesting properties of our class of states, many interesting questions are so far open. Is it possible to find a quasi local parent Hamiltonian for such states? Can the transfer matrix $\mathds{T}$ used to compute correlation functions in the operator representation be put in canonical form? Are there important \cbl{gauge transformations} our discussion in \ref{subsec:gaugeinvariance} ignores? \cbl{How do $V$ and $\alpha$ encode topological order and (local and global) gauge symmetries?} Can this approach be combined with techniques developed on the lattice to study Gauge theories with tensor networks~\cite{rico2014,tagliacozzo2014,haegeman2015,zohar2015}? Are there non-trivial non-Gaussian cTNS for which correlation functions can be computed exactly? Do (possibly regularized) cTNS generically obey the area law like their discrete counterparts? \cbl{Can cTNS be used to construct interesting toy models of the AdS/CFT correspondence?} To what extent does the bond field dimension $D$ quantify entanglement for (possibly only some) classes of cTNS?

\cbl{Tackling these questions is an important goal for future work, to fully extend the success of tensor networks from the lattice to the continuum.}

\begin{acknowledgments}
We are grateful to Denis Bernard, Adri\'an Franco-Rubio, Giacomo Giudice, Anne Nielsen, German Sierra, Guifr\'e Vidal, and Erez Zohar for helpful discussions. We thank two anonymous referees for valuable suggestions and comments. AT was supported by the Alexander von Humboldt foundation and the Agence Nationale de la Recherche (ANR) contract ANR-14-CE25-0003-01. \cbl{The research of JIC is partially supported by the ERC Advanced Grant QENOCOBA under the EU Horizon 2020 program (grant agreement 742102).}
\end{acknowledgments}

\bibliography{main}

\end{document}